\documentclass[aps,preprintnumbers,nofootinbibt,twocolumn, keywords]{revtex4}
\usepackage{graphicx}
\usepackage{amsmath}
\usepackage{bm}
\usepackage{color}

\def\be{\begin{equation}}
\def\ee{\end{equation}}
\def\bea{\begin{eqnarray}}
\def\eea{\end{eqnarray}}

\begin{document}

\title{Generalized Langevin equation with colored noise description of the stochastic oscillations of accretion disks}
\author{Tiberiu Harko$^1$}
\email{t.harko@ucl.ac.uk}
\author{Chun Sing Leung$^{2}$}
\email{csinleung@polyu.edu.hk}
\author{Gabriela Mocanu$^{3}$}
\email{gabriela.mocanu@ubbcluj.ro}
\affiliation{$^1$Department of Mathematics, University College London, Gower Street,
London WC1E 6BT, United Kingdom}
\affiliation{$^2$Department of Applied Mathematics, Polytechnic University, Hong Kong, P. R. China}
\affiliation{$^3$Faculty of Physics,
 Babes-Bolyai University, Cluj-Napoca, Romania},

\date{\today}

\begin{abstract}
We consider a description of the stochastic oscillations of the general relativistic accretion disks around compact astrophysical objects interacting with their external medium  based on a generalized Langevin equation with colored noise, which accounts for the general memory and retarded effects of the frictional force, and on the  fluctuation-dissipation theorems. The presence of the memory effects influences the response of the disk to external random interactions, and modifies the dynamical behavior of the disk, as well as the energy dissipation processes. The generalized Langevin equation of the motion of the disk in the vertical direction is studied numerically, and  the vertical displacements, velocities and luminosities of the stochastically perturbed disks are explicitly obtained for both the Schwarzschild and the Kerr cases. The Power Spectral Distribution (PSD) of the disk luminosity is also obtained. As a possible astrophysical application of the formalism we investigate the possibility that the Intra Day Variability (IDV) of the Active Galactic Nuclei (AGN) may be due to the stochastic  disk instabilities.  The perturbations due to colored/nontrivially correlated noise induce a complicated disk dynamics, which could explain some astrophysical observational features related to disk variability.
\end{abstract}

\maketitle


\section{Introduction\label{intro}}

In an astrophysical environment, compact general relativistic objects like black holes or neutron stars are
often surrounded by an accretion disk. The physics of accretion disks around compact objects is thought to explain a number of observations related to X-ray binaries or Active Galactic Nuclei (AGN). The disks are usually considered as being composed from massive test particles that move in the gravitational field of the central compact object. Numerical simulations have shown that the accretion induced collapse  of a white dwarf may lead to a neutron star  surrounded by a disk with mass up to $0.8M_{\odot}$ \cite{Abd}.  Waves and normal-mode oscillations in geometrically thin and thick disks around compact objects have been studied extensively both within Newtonian gravity,  and within a relativistic framework (see \cite{Katoa} for a review of accretion disk properties). It was also suggested that the Intra Day Variability (IDV) of AGNs can also be related to the vertical oscillations of the accretion disks around black holes \cite{Min, Leung}.

Stochastic processes and methods play a fundamental role in our understanding of many physical/astrophysical phenomena \cite{Ch1, Chan, Frid}. The study of the general relativistic oscillations of thin accretion disks around compact astrophysical objects interacting with the surrounding medium through non-gravitational forces was considered in \cite{Harko}. The interaction of the accretion disk with the external medium  was
modelled via a friction force and a random force, respectively. The general equations describing the stochastically perturbed disks were derived by considering the perturbations of trajectories of the test particles in equatorial orbits, assumed to move along the geodesic lines. By taking into account the presence of a viscous dissipation and of a stochastic force the dynamics of the stochastically perturbed disks can be formulated in terms of a general relativistic Langevin equation. The stochastic energy transport equation was also obtained. The vertical oscillations of the disks in the
Schwarzschild and Kerr geometries were considered in detail, and the vertical displacements, velocities and luminosities of the stochastically perturbed disks were explicitly obtained for both the Schwarzschild and the Kerr cases by numerically integrating the corresponding Langevin equations.

The Langevin equation, used to describe the stochastic oscillations of the accretion disks in \cite{Harko}, provides a correct phenomenological and statistical description of the Brownian motion only in the long time limit, for times long as compared to the characteristic relaxation time of the velocity autocorrelation  function \cite{Co04, Boon}. In order to describe the dynamics of a homogeneous system without restriction on a time scale, one should generalize the Langevin equation by introducing, instead of the simple friction term, a systematic force term with an integral kernel~\cite{ Boon,Co04}. The convolution term expresses the {\it memory}, or {\it retardation} effect, which can make the accretion disk response to an external perturbing stochastic force quite complicated. As shown in~\cite{Harko}, a simple Brownian motion model for the disk oscillations produces a Power Spectrum Distribution (PSD) $P(f) \sim f^{-\alpha}$, with $\alpha =-2$.
Completely uncorrelated interaction of the disk with the external medium would produce white
noise, with a PSD $P(f) = f^0 = {\rm const}$. However, the observational data for a large percentage of AGN show that their optical IDV has PSDs for which $\alpha$ is neither $0$ nor $-2$, as shown by discrete Fourier transform analysis~\cite{Azarnia}, structure function analysis~\cite{Carini} and fractal dimension analysis~\cite{Leung2011}. Therefore, if this type of variability is produced and has its origin in the accretion disks rotating around compact astrophysical objects, a consistent mathematical description of the stochastic phenomena in accretion disks would require to go beyond the simple description of the complex accretion disk - cosmic environment interaction in terms of a simple friction and a stochastic force, respectively. The Langevin type equation with a damping term and stochastic force was used in \cite{SR0} to describe the stochastic oscillations on the vertical direction of the accretion disks around a black hole, and to calculate the luminosity and PSD for the oscillating disk. The stochastic resonance  phenomenon in PSD curves for different parameter values of viscosity coefficient, accretion rate, mass of black hole and outer radius of the disk was also studied. The results show that the simulated PSD curves of luminosity for disk oscillation have the same profile as the observed PSD of black hole X-ray binaries, and the stochastic resonance of accretion disk oscillation may be an alternative interpretation of the persistent low-frequency quasi-periodic oscillations.

A description of the stochastic oscillations of the general relativistic accretion disks around compact astrophysical objects based on the generalized Langevin equation, which accounts for the general retarded effects of the frictional force, and on the fluctuation-dissipation theorems, was introduced in \cite{GL}. The vertical displacements, velocities and luminosities of the stochastically perturbed disks were explicitly obtained for both the Schwarzschild and the Kerr cases. The PSD of the simulated light curves was determined and it was found that the spectral slope $\alpha$ has values that correspond with observations. The theoretical predictions of the model were compared with the observational data for the luminosity time variation of the BL Lac S5 0716+714 object.
The influences of the friction parameter, spin parameter  and mass $M$ of the central compact object on the stochastic resonance  in PSD curves was  discussed, by using the generalized Langevin equation, in \cite{SR1}. The results show that a large spin parameter  can enhance the stochastic resonance phenomenon, but the larger the friction coefficient or the central mass is, the weaker the stochastic resonance phenomenon becomes. The simulated PSD curves of the output luminosity of stochastically oscillating disks have the same profile as the observed PSDs of X-ray binaries. Hence the resonance peak in the PSD curve can explain the quasi-periodic oscillations.

It is the purpose of the present paper to consider the memory/retardation effects  in the description of the stochastic oscillations of the particles in the accretion disks around compact objects  under the influence of the external environment, generating some random interactions between  the disk and the cosmic medium. More specifically, we consider two astrophysical accretion disk - external medium interaction models that can be described by means of the Langevin type stochastic differential equations. If the black hole - accretion disk system is located in a dense cosmic environment, like, for example, a stellar cluster, then the gravitational force exerted by the neighbouring stars can be described as an effective stochastic, time dependent force \cite{Ch1, Chan}. The stochastic gravitational force appears due to the local fluctuations of the numbers of stars in the clusters, as shown in \cite{Ch1,Chan}, where these processes have been studied in much detail. Therefore for a massive central object - accretion disk system  located in a dense stellar cluster, the accretion disk is subjected to two gravitational forces, one produced by the black hole, and a second one representing an effective random force due to the neighboring stars. In this case the dynamical evolution of the accretion disk can be described by means of the Langevin equation.

As a second astrophysical process that can be modelled by a Langevin type stochastic differential equation we consider the dynamical  interaction between the accretion disk  and the remnant matter from a star tidally disrupted by the central black hole. In this case, by assuming that the mass of the disk is much smaller than the mass of the inflowing gas, one can model the system as a single particle (the disk) located into the hydrodynamical flow of the remnant gas. Under these simplifying assumptions the dynamics of the system can be described again by a stochastic Langevin type equation.

Moreover, in the present analysis we assume that the stochastic perturbations  of the disk (the particle), assumed to randomly interact with an external medium (fluctuating gravitational force and gas remnants of a tidally disrupted star), are described by the generalized Langevin equation, introduced in~\cite{Kubo}.

Mathematically, the stochastic equation of motion of the disk becomes a differential - integral equation, whose integral kernel describes the memory effects. We use this equation to provide a detailed description of the vertical stochastic oscillations of an accretion disk around a Schwarzschild black hole and also of an accretion disk around a Kerr black hole. The solution of the generalized Langevin equation is obtained numerically, and the basic physical parameters of the disk (vertical displacement, velocity and luminosity) are obtained in a numerical form. As a possible application of the present formalism we investigate the possibility that the IDV of the AGNs has its origin in some random/stochastic effects at the accretion disk level. In the present study we will focus only on supermassive black holes, both static and rotating, which can be described by the Schwarzschild and Kerr geometries, and which are the best candidates to explain the electromagnetic radiation emission properties of the AGNs,  assumed to be powered by accretion of mass onto massive black holes, with masses of the order of $10^6M_{\odot}$ to $10^{10}M_{\odot}$.

The present paper is organized as follows. The generalized Langevin equation, describing the stochastic oscillations of the accretion disks with memory is presented in Section~\ref{sect2}. Two specific examples of astrophysical interest of disk - external medium interaction that can be described by a Langevin type equation are also considered.  The dimensionless form of the disk equation of motion, as well as the numerical procedure for solving the generalized Langevin equation are presented in Section~\ref{sect3}. The stochastic oscillations of the accretion disks in the presence of colored noise are studied for both Schwarzschild and Kerr geometries in Section~\ref{sect4}. The application of the present formalism for the study of the IDV's is considered in Section~\ref{sect5}. We conclude and discuss our results in Section~\ref{sect6}.

\section{The generalized Langevin equation for stochastic oscillations of accretion disks}\label{sect2}

In the present Section we consider the equation of motion of the stochastically oscillating accretion disks interacting with their external cosmic environment.  We consider two specific cases of astrophysical interest, namely, the stochastic perturbations of the accretion disks, located in dense stellar clusters, by the gravitational fields of their neighboring stars, and the interaction between the accretion disk with the gaseous hydrodynamic flow generated by the disruption of a  star by the tidal forces of the central supermassive black hole.  By assuming that the particles in the disk move on geodesic lines, we show first that the effects of a stochastic perturbation can be described, in the Newtonian approximation, by a Langevin type equation. The Langevin equation with memory is introduced, and the connection between the generalized Langevin equation and the fluctuating hydrodynamics \cite{Lali,Boon} is also established.

\subsection{Equation of motion of massive test particles in an accretion disk}

 We model the accretion disk as a single thermodynamic system, consisting of a large number of particles,  with overall energy density $\epsilon $ and pressure $p$, respectively. The energy-momentum tensor $T_{\mu \nu }$ of the disk is given by
\be
T_{\mu \nu}=\left(\epsilon +p\right)u_{\mu }u_{\nu }-pg_{\mu \nu},
\ee
where the four-velocity $u_{\mu }$ of the particles with mass $m$ satisfies the normalization condition $u_{\mu }u^{\mu }=1$, and the relation $u^{\mu }\nabla _{\nu }u_{\mu }=0$, respectively, where $\nabla _{\nu }$ represents the covariant derivative with respect to the metric. We also introduce the projection operator $\overline{h}_{\mu \lambda } = g_{\mu \lambda } - u_{\mu }u_{\lambda }$, which satisfies the relation $\overline{h}_{\mu \lambda }u^{\mu } = 0$. The equations of motion of the particles in the disk follow from the equation of conservation of the energy-momentum tensor, $\nabla _{\mu }T_{\nu }^{\mu }=0$, and, after contraction with $\overline{h}_{\mu \lambda }u^{\mu }$, leads to the equations
\be
\left(\epsilon +p\right)g_{\mu \lambda }u^{\nu }\nabla _{\nu }u^{\mu }-\nabla _{\nu }\overline{h}^{\nu }_{\lambda }=0.
\ee
Finally, contraction with $g^{\alpha \lambda }$ gives the equation of motion of the fluid element of the disk as  \cite{Klei}
\be
\frac{d^{2}x^{\mu }}{ds^{2}}+\Gamma _{\alpha \beta }^{\mu }\frac{dx^{\alpha }%
}{ds}\frac{dx^{\beta }}{ds}=\frac{p_{,\nu }}{\epsilon +p}\overline{h}^{\alpha \nu },
\end{equation}
where $\Gamma _{\alpha \beta }^{\mu }$ are the Christoffel symbols
associated to the metric, and a comma denotes the partial derivative with respect to the coordinate $x^{\nu }$.

We consider that the particles in the disk are in contact with
the external environment (stochastic gravitational field of neighboring compact astrophysical objects, remnant gas from tidally disrupted stars flowing towards the central black hole etc.), whose action on the disk can be  described generally by means of a stochastic force $\xi^{\mu }$.   If the disk is perturbed, a point particle in the
nearby position $x^{\prime \mu }$  must  satisfy the  generalized
equation of motion
\begin{equation}
\frac{d^{2}x^{\prime \mu }}{ds^{2}}+\Gamma _{\alpha \beta }^{\prime \mu }\frac{dx^{\prime \alpha }%
}{ds}\frac{dx^{\prime \beta }}{ds}=\frac{p^{\prime }_{,\nu }}{\epsilon +p}\overline{h}^{\prime \alpha \nu }+\frac{1}{m}%
\xi ^{\mu },
\end{equation}
where $\xi ^{\mu }$ is the external, random force, acting on the
particle. The coordinates $x^{\prime \mu }$ are given by $x^{\prime \mu
}=x^{\mu }+\delta x^{\mu }$, where $\delta x^{\mu }$ is a small quantity.
Thus, in the first approximation, we obtain for the Christoffel symbols
\begin{equation}
\Gamma _{\alpha \beta }^{\prime \mu }\left( x^{\sigma }+\delta x^{\sigma }\right) =\Gamma
_{\alpha \beta }^{\mu }\left( x^{\sigma }\right) +\Gamma _{\alpha \beta ,\lambda
}^{\mu }\left( x^{\sigma }\right)\delta x^{\lambda },
\end{equation}
where $\Gamma _{\alpha \beta ,\lambda }^{\mu }=\partial \Gamma _{\alpha
\beta }^{\mu }/\partial x^{\lambda }$. Moreover,
\be
u^{\prime }\left(x^{\prime \mu }\right)=u^{\mu }\left(x^{\mu }\right)+\frac{d\delta x^{\mu }}{ds},
\ee
\be
g^{\prime \mu \nu}\left(x^{\prime \alpha }\right)=g^{\mu \nu }\left(x^{\alpha }\right)+\nabla ^{\nu }\delta x^{\mu }+\nabla ^{\mu }\delta x^{\nu },
\ee
and
\be
p^{\prime }_{,\nu }\left(x^{\prime \alpha }\right)=p_{,\nu }\left(x^{\alpha }\right)+p_{,\mu ,\lambda }\left(x^{\alpha }\right)\delta x^{\lambda },
\ee
respectively. In the following we introduce the four-velocity of the perturbed motion as $ V^{\mu }=d\delta x^{\mu }/ds$. $V^{\mu }$ satisfies the condition $u_{\mu}V^{\mu }=0$.

Therefore the equation of motion for $\delta x^{\mu }$ becomes
\bea\label{finf}
&&\frac{d^{2}\delta x^{\mu }}{ds^{2}}+2\Gamma _{\alpha \beta }^{\mu }u^{\alpha }V^{\beta }+\left(\Gamma _{\alpha \beta ,\lambda
}^{\mu }u^{\alpha }u^{\beta }-\frac{p_{,\nu ,\lambda }\overline{h}^{\alpha \mu }}{\epsilon +p}\right)\delta x^{\lambda
}=\nonumber\\
&& -\frac{p_{,\nu}}{\epsilon +p}\left(\nabla ^{\nu }\delta x^{\mu }+\nabla ^{\mu }\delta x^{\nu }+u^{\mu }V^{\nu }+u^{\nu }V^{\mu }\right)+\xi ^{\mu }.
\eea

In obtaining Eq.~(\ref{finf}) we have assumed that the matter content of the disk can be modelled as a perfect fluid, and we have neglected the possibility of energy dissipation, either by viscous processes, or by radiation emission, considered to have a negligible effect on the disk structure and dynamics.  Moreover, we have adopted a full general relativistic approach, which takes into effect the modifications of the space-time geometry in the disk due to the presence of the central supermassive black hole. For realistic astrophysical systems, the study of Eq.~(\ref{finf}) can be done by using numerical methods only. In the following Sections we will obtain some approximate, but still realistic,  forms of Eq.~(\ref{finf}), which allow a simpler numerical treatment of the problem.

\subsection{The generalized Langevin equation}

Since the self-gravity of the disk is negligibly small as compared to the gravitational field of the central supermassive black hole, the infinitesimal transformations $x^{\prime \mu
}=x^{\mu }+\delta x^{\mu }$ do not change the metric. Therefore the coordinate transformations satisfy the Killing equations  $\nabla ^{\nu }\delta x^{\mu }+\nabla ^{\mu }\delta x^{\nu }=0$.

Hence it follows that in this approximation the equation of motion of the disk particles perturbed by an external force is given by
\bea\label{fin4}
&&\frac{d^{2}\delta x^{\mu }}{ds^{2}}+2\Gamma _{\alpha \beta }^{\mu }u^{\alpha }V^{\beta }+\left(\Gamma _{\alpha \beta ,\lambda
}^{\mu }u^{\alpha }u^{\beta }-\frac{p_{,\nu ,\lambda }\overline{h}^{\alpha \mu }}{\epsilon +p}\right)\delta x^{\lambda
}=\nonumber\\
&& -\frac{p_{,\nu}}{\epsilon +p}\left(u^{\mu }V^{\nu }+u^{\nu }V^{\mu }\right)+\xi ^{\mu }.
\eea
In the case of a disk consisting of particles rotating with azimuthal
angular velocity $\Omega =d\phi /dt$, the components of the four velocity are given by
$u^{\mu }=u^{t}\left( 1,0,\Omega /c,0\right)$.

If the disk consists of dust particles, with negligible pressure,  moving along the geodesic lines, the term containing the pressure gradient can be neglected, and we obtain the equation of motion considered in \cite{Shi}. In the adopted approximation of non-zero pressure, the deviation equation contains some new terms, which are proportional to the ratio of the second derivative of the pressure and the sum of the energy density and pressure, respectively, and the product of the pressure gradient and the deviation velocity $V^{\mu }$, respectively.

In the non-relativistic case the friction force is given by $%
f_{fr}^{i}=-\nu m v^{i}$, where $\nu $ is the friction coefficient, and $%
v^{i} $ are the components of the non-relativistic velocity \cite{Co04}. The
relativistic generalization of the friction force requires the introduction
of the friction tensor $\nu _{\alpha }^{\mu }$, given in \cite{Du05a, Du05b, Du09} as
\begin{equation}
\nu _{\alpha }^{\mu }=\nu m\left( \delta _{\alpha }^{\mu }+V_{\alpha }V^{\mu
}\right) .
\end{equation}

Therefore the friction force can be expressed  as \cite{Du05a, Du05b, Du09}
\begin{equation}
f_{fr}^{\mu }=-\nu _{\alpha }^{\mu }\left( V^{\alpha }-u^{\alpha }\right)
=-\nu m\left( \delta _{\alpha }^{\mu }+V_{\alpha }V^{\mu }\right) \left(
V^{\alpha }-u^{\alpha }\right) .
\end{equation}

In the following we will consider only the vertical oscillations of the
disk, which for an arbitrary axisymmetric metric are described by the equation
\begin{equation}
\frac{d^{2}\delta z}{ds^{2}}+\left[\omega _{\perp}^2  \left(
u^{t}\right) ^{2}-\frac{p_{,z,z}}{\epsilon +p}g^{zz}\right]\delta z=-\nu \left( \delta _{\alpha }^{z}+V_{\alpha
}V^{z}\right) V^{\alpha }+\xi ^{z},
\ee
where
\be
\omega _{\perp}^2=\Gamma _{tt,z}^{z}+2\Gamma _{t\phi
,z}^{z}\frac{\Omega }{c}+\Gamma _{\phi \phi ,z}^{z}\left(\frac{\Omega }{c}\right)^{2},
\ee
is the frequency of the gravitational oscillations of the disk \cite{Harko}.

In the following we assume that the gravitational effects, which enter in Eq.~(\ref{finf}) via the derivatives of the Christoffel symbols $\Gamma _{\alpha \beta ,\lambda }^{\mu }$  are much stronger than the hydrodynamic pressure effects, and therefore $\omega _{\perp}^2  \left(
u^{t}\right) ^{2}>>p_{,z,z}g^{zz}/\left(\epsilon +p\right)$.

By considering the non-relativistic limit of Eq.~(\ref{fin4}), the perturbed stochastic geodesic equation describing the vertical oscillations of the disk is equivalent with the Langevin type equation \cite{Harko}
\begin{equation}\label{L1}
\frac{d^2\delta z}{dt^2}+c\nu \frac{d\delta z }{dt}+c^2\omega _{\perp}^2\delta z=c^2\xi ^z(t).
\end{equation}

Eq.~(\ref{L1})  practically represents a linear oscillator, with a viscosity term, and a random perturbing force. In obtaining Eq.~(\ref{L1}) we have assumed that the external perturbative force is a function of time only, and we have neglected
its possible spatial coordinate dependence. A fundamental assumption of non-relativistic Galilean physics is the existence of a universal time $t$. Therefore, within the non-relativistic Langevin theory, it is quite natural to identify this universal time $t$ with the time parameter of the stochastic driving process \cite{Du09}. In general, the external stochastic noise $\xi ^z$ can be represented as a function of a $t$-parameterized two dimensional stochastic process, $\xi ^z=\xi ^z\left[x^{\mu }(t),p^{\mu}(t)\right]$, where $p^{\mu}$ is the four-momentum of the external source. By assuming that the motion is non-relativistic, with $v/c<<1$, and $\left|\vec{p}\right|<< mc$, where $\left|\vec{p}\right|$ is the absolute value of the momentum, we have $E\approx mc^2\approx {\rm constant}$, and consequently the dependence on the four-momentum of the stochastic force can be neglected. For a spatially inhomogeneous external perturbing cosmic environment, the function $\xi ^z$ would also depend on $z$. However, if the geometric dimension of the external stochastic source is much larger than the size of the disk, the inhomogeneities in the generated stochastic force can be neglected.  Hence, in the following the external noise $\xi ^z$  is  modeled by the 1-dimensional standard homogeneous Wiener process, with any possible momentum or space-like coordinate dependence neglected.

The Langevin equation Eq.~(\ref{L1}) provides a correct phenomenological and statistical description of the Brownian motion only in the long time limit, for times long as compared to the characteristic relaxation time of the velocity autocorrelation  function \cite{Co04, Boon}. In order to describe the dynamics of a homogeneous system without restriction on a time scale, one should generalize the Langevin equation by introducing, instead of the simple friction term, a systematic force term with an integral kernel~\cite{ Boon,Co04}. The convolution term expresses the {\it memory}, or the {\it retardation} effects, and, in the specific case of the stochastic oscillations of the accretion disks, the effects of the non-zero disk pressure and energy density.

In the case of the vertical oscillations of a thin accretion disk, the generalized Langevin equation, introduced in~\cite{Kubo}, which takes into account the retarded effects of the frictional force, is given by
\begin{equation}\label{genLang}
\frac{d^2\delta z}{dt^2}+\int_0^t{\gamma (t-\tau) \frac{d\delta z (\tau )}{d\tau}}d\tau+ K'(\delta z)=c^2 \xi ^z(t),
\end{equation}
where
\begin{equation}\label{gamma}
\gamma (\vert t-\tau \vert) = \frac{\alpha _1c}{\tau _d} \exp \left \{ -\frac{\vert t-\tau \vert}{\tau _d} \right \},
\end{equation}
and
\begin{equation}
K (\delta z) = \frac{c^2 \omega _\perp ^2}{2} \left ( \delta z \right)^2,
\end{equation}
respectively. In Eq.~(\ref{gamma}) $\tau _d$ represents a characteristic disk memory time, while $\alpha _1$ is interpreted as a disk damping strength. For $\vert t-\tau \vert<<\tau _d$,$\gamma (\vert t-\tau \vert)\approx \alpha _1\times c/ \tau_d$, while for $\vert t-\tau \vert>>\tau _d$,  $\gamma (\vert t-\tau \vert)\approx 0$. From a physical point of view the coefficient $\alpha _1$ describes the strength of energy dissipation in the disk. The autocorrelation function of the stochastic force satisfies the condition
\begin{equation}\label{mean}
\langle \xi ^z (t) \xi ^z (\tau) \rangle = \frac{1}{\beta} \gamma (t-\tau),
\end{equation}
where $\beta $ is a constant.

It is worthwhile to summarize now the physical assumptions, implicitly assumed by Eqs.~(\ref{genLang}), (\ref{gamma}), and (\ref{mean}):

a) The external cosmic environment interacting with the accretion disk around a supermassive black hole is spatially homogeneous and stationary; i.e., relaxation processes within the external environment occur on time scales much shorter than the relevant dynamical time scales associated with the motion of the accretion disk. Interaction with a spatially inhomogeneous non-stationary cosmic environment can be modelled  by considering friction and noise amplitude functions depending on the spatial coordinates, and disk momenta.

b) On a macroscopic level, the interaction between the disk particles and
the external cosmic environment is sufficiently well described by the generalized friction coefficient with memory $\gamma (\vert t-\tau \vert)$, and the
stochastic Langevin force $\xi ^z(t)$.

c) Stochastic collisions between the disk particles and the constituents of the external environment occur virtually uncorrelated.

In the next Section we will use Eq.~(\ref{genLang}) to study the vertical oscillations of the accretion disks around supermassive black holes in both Schwarzschild and Kerr geometries.

\subsection{Equation of motion of a thin disk in the presence of a stochastic gravitational field}

 As a first example of the stochastic interaction between accretion disks around supermassive black holes and the cosmic environment we consider the situation in which the disk, and its central object, is a member of a larger stellar system, in which all components interact gravitationally.  We assume that at the center of the disk we have a
compact object of mass $M$. The force acting on the disk can be divided into two components, the gravitational force of the central object, which we denote by $\vec{F}_G$, and the force $\vec{F}_R$ generated by the influence of the immediate local neighborhood, which may consist of ordinary stars and black holes. The force $\vec{F}_G$ is a smoothly varying
function of position, while $\vec{F}_R$ can be subject to relatively rapid fluctuations \cite{Ch1, Chan}. The fluctuation in the external force
is a direct consequence of the changing of the local stellar distribution that makes the influence of the near neighbors on the disk time
variable. It can be shown \cite{Ch1,Chan} that the fluctuations in the force acting on the central object  disk system, due to the changing local stellar distribution, do occur with extreme rapidity as compared to the rate at which any of the other physical parameters change. Let $p$ be the pressure of the matter in the disk. Therefore we can write for the
force acting on the disk the expression $\vec{F}=\vec{F_G}+\vec{F_p}+\vec{F}_{visc}+\vec{F}_R(t)$, where $\vec{F}_p$ is the force due to the pressure variation in the disk, $\vec{F}_{visc}$ is the force generated by the viscous dissipative effects in the disk, and the fluctuating random force $\vec{F}_R$, due to the gravitational interaction of the disk with the nearby stellar systems, can be considered as a function of time only \cite{Ch1, Chan}.

In order to obtain the equation of motion of the disk in the presence of the gravitational force of the central object, and of the external stochastic forces, we consider that the disk is thin in the sense discussed in~\cite{ShSu,ShSu2}. We approximate the distribution of the surface
density $\Sigma $  by the expressions \cite{Tit}
\begin{equation}\label{d1}
\Sigma (r)=\left\{
\begin{array}{lll}
\Sigma _{0}={\rm constant},R_{in}\leq r\leq R_{adj}, & & \\
&  &  \\
\Sigma _{0}\left( \frac{r}{R_{adj}}\right) ^{-\chi },R_{adj}\leq
r\leq R_{out}, &  &  \\
&  &  \\

\end{array}%
\right.
\end{equation}
where $R_{in}$ is the innermost radius of the disk, $R_{adj}$ is an
adjustment radius in the disk, and $R_{out}$ is the outer radius of the
disk. The index $\chi $ of the surface density can be either $3/5$ or $3/4$  \cite{Tit}.

The mass of the disk $M_{d}$ from $R_{in}$ to $R_{out}$ is given by
\begin{eqnarray}
M_{d}&=&2\pi \int_{R_{in}}^{R_{out}}\Sigma \left( r\right) rdr\approx \frac{%
2\pi \Sigma _{0}R_{adj}^{2}}{2-\chi }\left( \frac{R_{out}}{R_{adj}}\right)
^{2-\chi },\nonumber\\
&&\chi \neq 2, z<<R_{in},
\end{eqnarray}
where we have used  the assumptions $R_{in}<<R_{out}$ and $R_{adj}\approx R_{out}$, respectively. We assume that the disk as a whole is
perturbed, and that it deviates from the equatorial plane by a small
distance $\delta z$. The restoring gravitational force $F_{G}$ caused by the
attraction of the central object is \cite{Tit}
\begin{equation}
F_{G}(\delta z)=GM\delta z\int_{R_{in}}^{R_{out}}\frac{2\pi r\Sigma \left( r\right) }{%
\left( r^{2}+z^{2}\right) ^{3/2}}dr.
\end{equation}

After integration, using the density distribution given by Eqs.~(\ref{d1}), we obtain
\begin{equation}
F_{G}(\delta z)=\frac{2\pi GM\Sigma _{0}}{R_{in}}\left( 1-\frac{\chi }{\chi +1}%
\frac{R_{in}}{R_{adj}}\right) \delta z.
\end{equation}

We define the mean vertical density of the disk as $\rho _d=M_d/h_0$, where $h_0$ is the half-thickness of the disk.
Therefore the vertical oscillations of the disk as a whole can be described
by the equation of motion
\begin{equation}
\frac{d^{2}\delta z}{dt^{2}}+\frac{1}{M_d}F_{G}(\delta z)+\frac{1}{\rho _d}\frac{\partial p}{\partial z}+F_{visc}=\frac{1}{M_d}F_R(t).
\end{equation}
As for the acceleration due to the vertical pressure distribution we can approximate it as $a_{press}=-\left(1/\rho _d\right)\partial p/\partial z\approx -\kappa _z(r)$, where the vertical epicyclic frequency $\kappa_ z(r)$ depends on the horizontal radius $r$ only. The simplest way to obtain such a relation is by considering a power series expansion of the pressure so that
\be
p(r,z)=p(r,0)+\left.\frac{\partial p(r,z)}{\partial z}\right|_{z=0}z+\frac{1}{2}\left.\frac{\partial ^2p(r,z)}{\partial z^2}\right|_{z=0}z^2+....
\ee
The constant term in the acceleration can be eliminated by a rescaling of the acceleration, and the epicyclic frequency can be obtained as $\kappa _z(r)=\left.\partial ^2p(r,z)/\partial z^2\right|_{z=0}/\rho _d$. Therefore for the equation of motion of the disk we obtain
\begin{equation}
\frac{d^{2}\delta z}{dt^{2}}+\left(\omega _{0}^{2}+\kappa _z^2\right)\delta z+F_{visc}=\frac{F_R(t)}{M_d},  \label{mot}
\end{equation}
where
\begin{equation}
\omega _{0}^{2}=4\pi ^{2}\nu _{0}^{2}=\frac{\left( 2-\chi \right) GM}{%
R_{in}R_{adj}^{2}}\left( \frac{R_{out}}{R_{adj}}\right) ^{\chi -2}\left( 1-%
\frac{\chi }{\chi +1}\frac{R_{in}}{R_{adj}}\right) .
\end{equation}
If some dissipative forces, like, for example, dissipation due to radiation emission, or viscosity friction effects in the disk matter, are also acting on the disk, one should include the dissipative forces in the equation of motion. The simplest phenomenological approach would be to assume that the viscous/heat dissipation is proportional to the velocity of the vertical oscillations, $F_{visc}\propto \delta z/dt$. However, a better phenomenological description of the viscous friction can be obtained by assuming an integral kernel description of the dissipative processes. Therefore the equation of motion of the disk becomes
\be
\frac{d^{2}\delta z}{dt^{2}}+\int_0^t{\gamma (t-\tau) \frac{d\delta z (\tau )}{d\tau}}d\tau+\left(\omega _{0}^{2}+\kappa _z^2\right)\delta z=\frac{F_R(t)}{M_d}.  \label{L2}
\ee
In this model the disk as a whole undergoes stochastic oscillations determined by the interaction between the disk, and the fluctuating stellar/black hole neighborhood.  The equation of motion is given by the generalized Langevin Eq.~(\ref{L2}).

In order to obtain the oscillations frequency of the disk  we assume that the innermost radius of the disk $R_{in}$ is located at $R_{in}=6GM/c^2$, corresponding to the innermost stable orbits in a general relativistic Schwarzschild geometry. We represent the adjustment radius of the disk $R_{adj}$ as $R_{adj}=f_1 R_{in}$ and the outer radius of the disk as $R_{out}=g_1 R_{in}$, where $f_1 >1$ and $g_1 >>1$ are constants. Then for the oscillation frequency of the disk we obtain
\begin{equation}\label{omega0}
\omega _0=\frac{c^3}{6f_1 GM}\left(\frac{g_1 }{f_1 }\right)^{\chi /2-1}\sqrt{\frac{2-\chi }{6}\left(1-\frac{\chi }{\chi +1}\frac{1}{f_1 }\right)},
\end{equation}
or
\bea
\omega _0^2&=&1.89\times 10^{-4}\times \left(\frac{M}{10^6M_{\odot}}\right)^{-2}\times
\frac{2-\chi }{f_1 ^2}\left(\frac{g_1 }{f_1 }\right)^{\chi -2}\times \nonumber\\
&&\left( 1-\frac{%
\chi }{\chi +1}\frac{1}{f_1 }\right)\;{\rm s}^{-2}.
\eea
The period of the free disk oscillations  is given by
\bea
T_0&=&5.171\times 10^{-2}f_1 \left(\frac{M}{10^6M_{\odot}}\right)\left( \frac{g_1
}{f_1 }\right) ^{1-\chi /2}\times \nonumber\\
&&\left[ \frac{2-\chi }{6}\left( 1-\frac{%
\chi }{\chi +1}\frac{1}{f_1 }\right) \right] ^{-1/2}\;{\rm s}.
\eea
In the present paper we will consider only the case $\omega _0^2>>\kappa _z^2$, and therefore our study will be limited to the consideration of the gravitational effects in the disk.

\subsection{Stochastic equation of motion of accretion disks in a fluctuating environment due to stellar capture}

 Close encounters between stars from a nuclear cluster and a massive black hole lead to the disruption of stellar bodies due to the gravitational tidal forces of the black holes,  and the subsequent accretion of the remnant gas \cite{Baum, Karas}. These astrophysical processes  are among the likely mechanisms for feeding black holes that are embedded in a dense stellar cluster. Stars on highly elongated orbits are susceptible to tidal disruption and hence they provide a natural source of material to replenish the inner disk. Such events can produce debris and lead to recurring episodes of enhanced accretion activity, with the infalling matter interacting both gravitationally and mechanically with the accretion disk.  Intermediate-mass black holes  have a high chance of capturing stars through tidal energy dissipation within a few core relaxation times \cite{Baum}. The interaction between the accretion disk and the remnant gas falling towards the central black hole can be considered as a random process, in which the particles of the disk interact with the particles of the tidally disrupted star.

In order to model the interaction between the disk and the remnant matter from the disrupted star we assume that the disk is thin, and its mass $M_d$ is much smaller than the mass $M_F$ falling towards the black hole, $M_d<<M_F$, condition valid at least for finite  time interval $0<t<t_F$, where $t_F$ is the in-falling time of $M_F$. Hence, if this condition is satisfied, one can obtain a simplified description of the disk - remnant stellar matter interaction by considering the disk as a single physical system randomly moving in a fluid represented by the matter accreted by the black hole. This simple physical description allows the description  of the dynamical evolution of the disk in terms of a Langevin  type equation, obtained from the fluctuating hydrodynamic equations of motion of the interacting particle - fluid system.

 Therefore in the following we consider the accretion disk as a  macroscopic system of mass $M_d$ in an
incompressible fluctuating fluid (the debris from stellar disruption). The position of a point on the surface of the
accretion disk at time $t$ is described through the position vector $\vec{r}_1(t)$. We denote by $\vec{r}$  the position vector of a point in the fluid and by $\vec{R}$  the position vector of the center of the accretion disk.  The motion of the fluid (the gaseous remnant of the star) in the presence of an exterior (gravitational) force $\vec{F}_{ext}$ is described by
the linearized stochastic Landau-Lifshitz equations of motion \cite{Lali, FoUh,BeMa, Nelkin},
\begin{equation}
\rho _m \frac{\partial }{\partial t}\vec{u}\left( \vec{r},t\right) =-\nabla
\cdot \stackrel{\leftrightarrow}{P}\left( \vec{r},t\right) +\vec{F}_{ext}\left( \vec{r},t\right),\left| \vec{r}-\vec{R}\left( t\right)
\right| >r_1,  \label{1n}
\end{equation}
 where $\stackrel{\leftrightarrow}{P}\left( \vec{r},t\right)$ is the total fluid stress tensor, with components
\begin{equation}
P_{ij}=p\delta _{ij}-\eta \left( \frac{\partial }{\partial r_{i}}u_{j}+%
\frac{\partial }{\partial r_{j}}u_{i}\right) +\Pi _{ij},
\end{equation}
where $p$ is the pressure in the fluid, $\eta $ is the shear viscosity
coefficient of the matter falling towards the black hole, and $\stackrel{\leftrightarrow}{\Pi}$ is the random stress tensor, with components $\Pi_{ij}$, which has the
following stochastic properties \cite{Lali, FoUh}
\begin{equation}
\left\langle\stackrel{\leftrightarrow}{\Pi}\left( \vec{r},t\right) \right\rangle =0,\left|
\vec{r}-\vec{R}\left( t\right) \right| >r_1,
\end{equation}
\begin{eqnarray}
\left\langle \Pi _{ij}\left( \vec{r},t\right) \Pi _{kl}\left( \vec{r}%
^{\prime },t^{\prime }\right) \right\rangle &=&2k_B\Theta\eta
\Bigg( \delta
_{ik}\delta _{jl}+\delta _{il}\delta _{jk}-\nonumber\\
&&\frac{2}{3}\delta _{ij}\delta
_{kl}\Bigg) \times
\delta \left( \vec{r}-\vec{r}^{\prime }\right)
\delta \left(
t-t^{\prime }\right) ,\nonumber\\
&& \left| \vec{r}-\vec{R}\left( t\right) \right| >r_1,
\end{eqnarray}
where the square brackets denote averages over an equilibrium ensemble, $k_B$ is the Boltzmann constant and $\Theta $ is the temperature of the external cosmic environment. We have also denoted $\nabla
\cdot \stackrel{\leftrightarrow}{P}\left( \vec{r},t\right)=\partial P_{ij}/\partial x_j$.  We assume that the accreted gas is incompressible, and therefore its velocity satisfies the condition
\begin{equation}
\nabla \cdot \vec{u}=0.
\end{equation}

The motion of the accretion disk, modelled as a single physical system, immersed in the fluid accreted by the central black hole in the
presence of an external gravitational force $\vec{F}_{ext}\left( \vec{r},t\right)$ is governed by the equation \cite{Lali, FoUh, BeMa, Nelkin, Boon}
\begin{equation}
m\frac{d\vec{v}(t)}{dt}=\vec{F}(t)+\vec{F}_{ext}(\left( \vec{r},t\right)=-\int_{S(t)}\stackrel{\leftrightarrow}{P}%
\left( \vec{r},t\right) \cdot \vec{n}dS+\vec{F}_{ext}(t),  \label{6n}
\end{equation}
where $\stackrel{\leftrightarrow}{P}\left( \vec{r},t\right) \cdot \vec{n}=P_{ij}n_{j}$, and $\vec{n%
}$ is the unit vector perpendicular to the thin accretion disk, and $S(t)$ is the surface
area of the accretion disk. Eqs.~(\ref{1n}) and (\ref{6n}) simply express the
conservation of the momentum. In the fully linearized scheme one may neglect
the time dependence of $\vec{R}(t)$ and $S(t)$ in these equations, and take
the origin of the coordinates in the center of the accretion disk, so that $\vec{R}%
(t)=0$. The stochastic boundary value problem associated with the motion of
the accretion disk of radius $\vec{r}_1(t)$ in the fluid accreted by the black hole is specified by the boundary
condition
\begin{equation}
\vec{v}\left( \vec{r}_1(t),t\right) =\vec{V}\left( t\right) ,
\end{equation}
where $\vec{r}_1(t)$ is the position vector of a point on the surface of the
accretion disk at time $t$ having the velocity $\vec{V}(t)$. $\vec{r}_1(t)$ is given
by $\vec{r}_1(t)=\vec{r}_1\left( 0\right) +\int_{0}^{t}\vec{V}\left( t^{\prime
}\right) dt^{\prime }$. The components of the force acting on the surface of the accretion disk due to the fluid falling towards the black hole  are
given by \cite{Lali, FoUh, BeMa, Nelkin, Boon}
\begin{eqnarray}
F_{i}(t)&=&-\int_{S(t)}\left[ p\delta _{ij}-\eta \left( \frac{\partial }{%
\partial r_{i}}u_{j}+\frac{\partial }{\partial r_{j}}u_{i}\right) \right]
n_{j}dS-\nonumber\\
&&\int_{S(t)}\Pi _{ij}n_{j}dS.
\end{eqnarray}

By estimating the perturbations of the velocities in the hydrodynamic flow of the gas remnant
due to the presence of the accretion disk it follows that the equation of motion of
the perturbed accretion disk can be written in the form of a generalized Langevin equation of
the form \cite{Nelkin}
\begin{equation}
M_d\frac{d\vec{v}(t)}{dt}=\vec{F}_{ext}(t)+m\int_{0}^{t}\gamma \left(
t-t^{\prime }\right) \vec{v}(t^{\prime })dt^{\prime }+\xi (t),
\end{equation}
where the function $\gamma \left( t-t^{\prime }\right) $ represents a
retarded effect of the frictional force and $\xi (t)$ is the random force
generated by the stochastic fluctuations in the fluid accreted by the black hole. Assuming that the
flow is unsteady, one should also take into account the so-called  retarded
viscous force effect, which  is due to an additional term to the Stokes
drag, related to the history of the particle acceleration. This additional
drag force, nowadays referred to as the Basset history force, has the form \cite{Boon}
\begin{equation}
\vec{F}_{B}=6\pi r_p^{2}\eta \int_{0}^{t}\left( t-t^{\prime }\right) ^{-1/2}%
\frac{d\vec{v}\left( t^{\prime }\right) }{dt^{\prime }}dt^{\prime },
\end{equation}
where $r_p$ is the particle radius. However, in the present paper we ignore the effects of the Basset history force on the accretion disk dynamics.

\section{The numerical approach to the generalized Langevin equation}\label{sect3}

In the present Section we write down the equation of motion of the stochastically oscillating accretion disks in a dimensionless form. The numerical procedure to solve the Langevin equation in the presence of a colored noise is also described. To characterize the luminosity of the stochastically oscillating disks from a global point of view we will use the Power Spectral Distribution (PSD), whose main properties are briefly reviewed.

\subsection{Dimensionless form of the generalized Langevin equation}

By introducing a set of dimensionless parameters $(\theta, \sigma,  \overline{q})$ defined as
\begin{equation}
t=\tau _d\theta , \tau =\tau _d \sigma ,   \delta z=\frac{c^2\tau _d^2}{M_D}  \overline{q};
\end{equation}
where $M_{D}$ is the mass of the disk, the equation of motion of the stochastically oscillating disk given by Eq.~(\ref{genLang}) can be written as
\begin{equation}
\frac{d^2\overline{q}}{d\theta ^2}+a\int_0^{\theta }{e ^{-\vert \theta-\sigma \vert} \frac{d\overline{q} (\sigma )}{d\sigma}d\sigma }+b \overline{\omega} _\perp ^2 \overline{q}=\overline{\xi} (\theta ),\label{eq:GLEnoDim}
\end{equation}
where we have denoted
\begin{equation}
a=\alpha _1c\tau _d,b=\frac{c^2\tau _d^2}{M^2},\xi ^z = \frac{1}{M_D}\overline{\xi}(\theta),
\end{equation}
and
\begin{equation}
\omega _\perp  M =\overline{\omega} _ \perp ,
\end{equation}
respectively, where $M$ is the mass of the central compact object.

For the luminosity calculation, the energy is the sum of kinetic plus potential energy
\begin{equation}
E = \frac{1}{2} \left ( \frac{d \delta z}{dt} \right )^2 + K(\delta z)
\end{equation}
and the output luminosity is the time variation of the energy
\begin{equation}
L = - \frac{dE}{dt} = - \frac{d \delta z}{dt} \left [ \frac{d^2 \delta z}{dt^2} + K'(\delta z) \right ].
\end{equation}

The luminosity, written in dimensionless form as
\begin{equation}
L = \frac{c^4 \tau _d}{M_d^2} \overline{L},
\end{equation}
has an evolution given by
\begin{equation}
\overline{L} = - \frac{d\overline{q}}{d\theta} \left ( \frac{d^2 \overline{q}}{d\theta ^2} + b \overline{\omega}_\perp ^2 \overline{q} \right ).
\end{equation}

\subsection{The numerical procedure}

To solve the dimensionless generalized Langevin Eq.~(\ref{eq:GLEnoDim}), the algorithms already developed in the literature, see e.g.~\cite{hersh}, were used to obtain a set of three differential equations as follows
\begin{equation}
\dot{\overline{q}} = \overline{p}, \label{eq:appq}
\end{equation}
\begin{equation}
\dot{\overline{p}} = - b \overline{\omega}_\perp ^2 \overline{q} + \overline{z}, \label{eq:appp}
\end{equation}
\begin{equation}
\dot{\overline{z}} = - \overline{z} - a \overline{p} + \Upsilon (\theta),\label{eq:appz}
\end{equation}
where the noise $\Upsilon$ has the property
\begin{equation}
\langle \Upsilon (\theta) \Upsilon (\sigma) \rangle = A \delta (\theta - \sigma),
\end{equation}
and the initial value of $\overline{z}$ is drawn from a distribution with second moment
\begin{equation}
\langle \overline{z}(0)^2\rangle = 2 A.
\end{equation}

The dimensionless form of the disk luminosity is given by
\begin{equation}
\overline{L} = - \overline{p} \overline{z}.
\end{equation}

The  set of the stochastic linear differential equations (\ref{eq:appq})-(\ref{eq:appz}) is solved by implementing the fourth order Runge-Kutta integrator developed in~\cite{hersh}. Time is discretized with timestep $h$. At each timestep $j+1$ the values of the variables in the set $\mathcal{T} _{j+1} = \{\overline{q},\overline{p},\overline{z}\}$ are calculated as
\begin{equation}
x_{j+1} = x_j^{det}(\mathcal{T}_j) + x_j^{rand} (\mathcal{T}_j)\label{eq:set},
\end{equation}
i.e., as the sum of a deterministic part and a random part, where both parts depend on the full values of all the parameters in the set at the previous timestep. The deterministic part follows by evolving the "normal" Runge-Kutta algorithm in time and has the form
\begin{equation}
\left( \begin{array}{c}
\overline{q}_{j+1}^{det} \\
\overline{p}_{j+1}^{det} \\
\overline{z}_{j+1}^{det} \\
\end{array} \right ) = D
\left( \begin{array}{c}
\overline{q}_j \\
\overline{p}_j \\
\overline{z}_j \\
\end{array} \right ),
\end{equation}
where $D$ is a matrix with components
\begin{equation}
D_{11} = - \frac{h^2}{2} b \overline{\omega} _\perp ^2 + \frac{\overline{\omega}_\perp ^2 b h^4}{24} \left ( b \overline{\omega} _\perp ^2 + a\right ),
\end{equation}
\begin{equation}
D_{12} = h - \frac{h^3 b \overline{\omega} _\perp ^2}{6} - \frac{h^3 a}{6} + \frac{h^4 a}{24},
\end{equation}
\begin{equation}
D_{13} = \frac{h^2}{2} - \frac{h^3}{6} - \frac{h^4}{24} \left ( b \overline{\omega} _\perp ^2 + a - 1 \right ),
\end{equation}
\begin{equation}
D_{21} = - h b \overline{\omega} _\perp ^2 + \frac{h^3 b \overline{\omega} _\perp ^2}{6} \left ( b \overline{\omega} _\perp ^2 + a \right ) - \frac{h^4 a b \overline{\omega} _\perp ^2}{24},
\end{equation}
\begin{equation}
D_{22} = -\frac{h^2}{2} \left ( b \overline{\omega} _\perp ^2 + a \right ) + \frac{h^3 a}{6} + \frac{h^4}{24} \left ( b^2 \overline{\omega} _\perp ^4 + 2ab\overline{\omega} _\perp ^2 + a^2 - a \right ),
\end{equation}
\begin{equation}
D_{23} = h - \frac{h^2}{2} + \frac{h^3}{6} \left ( 1 - b \overline{\omega} _\perp ^2 - a \right ) + \frac{h^4}{24} \left( b \overline{\omega} _\perp ^2 + 2a-1 \right),
\end{equation}
\begin{equation}
D_{31} = \frac{h^2ab\overline{\omega} _\perp ^2}{2} - \frac{h^3 ab \overline{\omega} _\perp ^2}{6} + \frac{h^4 a b \overline{\omega} _\perp ^2}{24} \left ( 1-b\overline{\omega} _\perp ^2 - a \right ),
\end{equation}
\begin{equation}
D_{32} = -ah + \frac{h^2}{2} - \frac{h^3 a}{6}\left ( 1-b\overline{\omega} _\perp ^2 - a \right ) + \frac{h^4a}{24} \left ( 1-b\overline{\omega} _\perp ^2 -2 a \right ),
\end{equation}
and
\begin{equation}
D_{33} = - h - \frac{h^2}{2} (1-a) + \frac{h^3}{6} (2a-1) + \frac{h^4}{24} \left ( ab\overline{\omega} _\perp ^2 +a^2 - 3a +1\right ),
\end{equation}
respectively.

The random part is calculated as detailed in Section II.B of~\cite{hersh}, and for the case studied here has the form
\begin{equation}
\left( \begin{array}{c}
\overline{q}_j^{rand} \\
\overline{p}_j^{rand} \\
\overline{z}_j^{rand} \\
\end{array} \right ) =
\left( \begin{array}{cccc}
0 & 0 & 1 & -1 \\
0 & 1 & -1 & - b \overline{\omega} _\perp ^2 - a + 1 \\
1 & -1 &  1-a & 2a-1
\end{array} \right)
\left( \begin{array}{c}
Z_1 \\
Z_2 \\
Z_3 \\
Z_4 \\
\end{array} \right ),
\end{equation}
where the $Z_i$'s are linear combinations of four independent Gaussian variables $ \{ a_1, a_2, a_3, a_4\} \in \mathcal{N} (0,h)$
\begin{equation}
Z_1 = a_1,
\end{equation}
\begin{equation}
Z_2 = h \left ( \frac{a_1}{2}  + \frac{ a_2}{2\sqrt{3}} \right ),
\end{equation}
\begin{equation}
Z_3 = h^2 \left ( \frac{a_1}{3!} + \frac{2 \sqrt{3} a_2}{4!} + \frac{a_3}{\sqrt{6!}}\right ),
\end{equation}
\begin{equation}
Z_4 = h^3 \left ( \frac{a_1}{4!} + \frac{a_2 \sqrt{3}}{40} + \frac{a_3}{24\sqrt{5}} + \frac{a_4}{120 \sqrt{7}} \right )\label{eq:Gaussian}.
\end{equation}

\subsection{The Power Spectral Distribution}

As a parameter allowing a global characterization of the stochastic behavior of the disk we use the slope of the Power Spectral
Distribution (PSD) of the physical parameters of the disk.  The numerical values of the slopes of the PSD provide insight
to the nature of the mechanism leading to the observed
variability. In statistical analysis, if $X$ is some fluctuating
stationary quantity, with mean $\mu _X$ and variance $\sigma _X ^2$, then an
autocorrelation function for  $X$ is defined as~\cite{Vaseghi,Larsen}
\begin{equation}
R_X(\tau) = \frac{\langle \left ( X_s - \mu _X \right ) \left (
X_{s+\tau} - \mu \right) \rangle}{\sigma _X^2},
\end{equation}
where $X_s$ is the values of $X$ measured at time $s$ and $\langle
\rangle$ denotes averaging over all values $s$. The PSD is defined based on the correlation function as~\cite{Vaseghi,Larsen}
\begin{equation}
P(f) = \int _{-\infty} ^{+\infty} R_X(\tau) e^{-\imath 2\pi f \tau}
d\tau
\end{equation}
and it is straightforward to see its importance in terms of the
"memory" of a given process. For example, if $X$ is the B band (optical)
magnitude of the disk, the slope of the PSD of a time series of
$X$ provides insight to the degree of correlation the underlying
physical process has with itself. The system needs additional
energy to fluctuate and this mechanism is historically best
explained for Brownian motion, in which case the energy is
thermal. As an example of the connection between the process generating the light curve and the observed spectral slope, a Brownian motion process would produce a PSD of the shape $P(f) \sim f^{-2}$ while a completely uncorrelated evolution of a system would produce white noise, i.e., a PSD of the shape $P(f) = f^0 = {\rm const}$. From this it is important to remember that the slope of the PSD can be regarded as an indicator of the type of underlying process generating the light curves. As already stated in Section~\ref{intro}, the observed IDV light curves have spectral slopes that do not fit in the simple models of Brownian motion or uncorrelated evolution. However, in the case of experimental curves, one cannot know before-hand if the process is purely stationary.

The experimental and simulated light curves are analysed in this paper with the software .R which produces spectral slopes in accordance with the algorithm described in~\cite{Vaughan}.

In order to trust an interpretation that the spectral slope represents a sign of correlation in the signal, the investigated signal should be stationary. By definition, a stationary stochastic process is a process whose properties are constant in time. More precisely, the mean value and the variance of such a signal are time independent and the autocorrelation of the signal at different times depends only on the time difference and not on the actual time at which the correlation is evaluated.






In our model, the perturbing noise is a stationary stochastic process, as can be easily seen because of its prescribed property Eq.~\eqref{eq:set}. To show that the resulting light curve $\bar{L} = - \bar{p} \bar{z}$ is stationary, one should look at Equations~\eqref{eq:set}-\eqref{eq:Gaussian}. The stochastic part of the signals involved are given by linear combinations of Gaussian random variables (the $Z_i$s), so their properties are stationary. We thus expect that our interpretation of the spectral slope is somewhat accurate.

\section{Colored stochastic oscillations of the general relativistic disks}\label{sect4}

In the present Section we will obtain the basic physical parameters (displacement, velocity, and luminosity), of the vertically oscillating accretion disks in both static Schwarzschild geometry, and in the rotating Kerr geometry, respectively.

\subsection{Colored stochastic disk oscillations for the Schwarzschild geometry}

The general form of a static axisymmetric metric can be represented as~\cite{Bini}
\begin{equation}
ds^{2}=-e^{2U}c^{2}dt^{2}+e^{-2U}\left[ e^{2\zeta }\left( d\rho
^{2}+dz^{2}\right) +\rho ^{2}d\phi ^{2}\right] .
\end{equation}

For the Schwarzschild solution the functions $U$ and $\zeta $ are given by
\begin{equation}
U=\frac{1}{2}\ln \frac{\sqrt{\rho ^{2}+\left( M+z\right) ^{2}}+\sqrt{\rho ^{2}+\left(
M-z\right) ^{2}}-2M}{\sqrt{\rho ^{2}+\left( M+z\right) ^{2}}+\sqrt{%
\rho ^{2}+\left( M-z\right) ^{2}}+2M},
\end{equation}
\begin{equation}
\zeta =\frac{1}{2}\ln \frac{\left[ \sqrt{\left( M+z\right) ^{2}+\rho ^{2}}+%
\sqrt{\left( M-z\right) ^{2}+\rho ^{2}}\right] ^{2}-4M^{2}}{4\sqrt{\left(
M+z\right) ^{2}+\rho ^{2}}\sqrt{\left( M-z\right) ^{2}+\rho ^{2}}},
\end{equation}
where $M$ is the mass of the central compact object in geometrical units \citep{Bini}. The radial frequency at infinity $\Omega _{\perp }^{2}$ of the particles in the disk, and the proper free oscillation frequency $\omega _{\perp }^{2}$ of the disk can be obtained as \cite{Harko, Kato, Sem}
\begin{equation}
\Omega _{\perp }^{2}=\frac{e^{4U-2\zeta }}{1-\rho U_{,\rho }}U_{,zz},
\end{equation}
and
\begin{equation}
\omega _{\perp }^{2}=\frac{e^{2U-2\zeta }}{1-2\rho U_{,\rho }}U_{,zz},
\label{freq}
\end{equation}
respectively.
In the case of the static Schwarzschild geometry, the free vertical oscillation frequency of general relativistic disks can be written as  \cite{Harko, Sem}

\begin{equation}\label{Schfr}
\omega _{\perp }^{2}\left(M,\rho \right)=\frac{M}{\rho ^2\sqrt{M^2+\rho ^2}-2M^2\left(M+\sqrt{M^2+\rho ^2}\right)},
\end{equation}
where $M$ is expressed in natural units as $M=GM/c^2$, and $\rho $ is the distance from the black hole center in the radial plan of the disk.

If we express $\rho $ in terms of the mass of the central object as $\rho = n M$, then
\begin{equation}
\overline{\omega} _{\perp }^{2}\left(n \right)=\frac{1}{\left[\sqrt{1+n^2}\left(n^2-2\right)-2\right]}.
\end{equation}

For $M=10^8M_\odot$, $n = 10$, $\overline{\omega}_\perp ^2 = 0.001$ and fixed values of the integration step $h=1$, dimensionless initial displacement $\overline{q}(0) = 0$, dimensionless velocity $\overline{p} (0) = 0.01$ and $b=4$, we analysed two cases. All trajectories are obtained by mediating over 1000 stochastic trajectories (for which the variable $\overline{z}_0$ is redrawn from its distribution).

First for fixed $A = 0.0015$ we varied $a$ and obtained the behaviour of the dimensionless displacement, velocity and luminosity.  Then for fixed $a=0.03$ we varied $A$ and obtained the behaviour of the same physical parameters. The results of the numerical simulations of the displacements, velocities and luminosities  for the stochastic oscillations of the accretion disks are presented in Figs.~\ref{fig:sch-q1}, \ref{fig:sch-p1}, and \ref{fig:sch-lum1}, respectively.

   \begin{figure*}
   \centering
  \includegraphics[width=8cm]{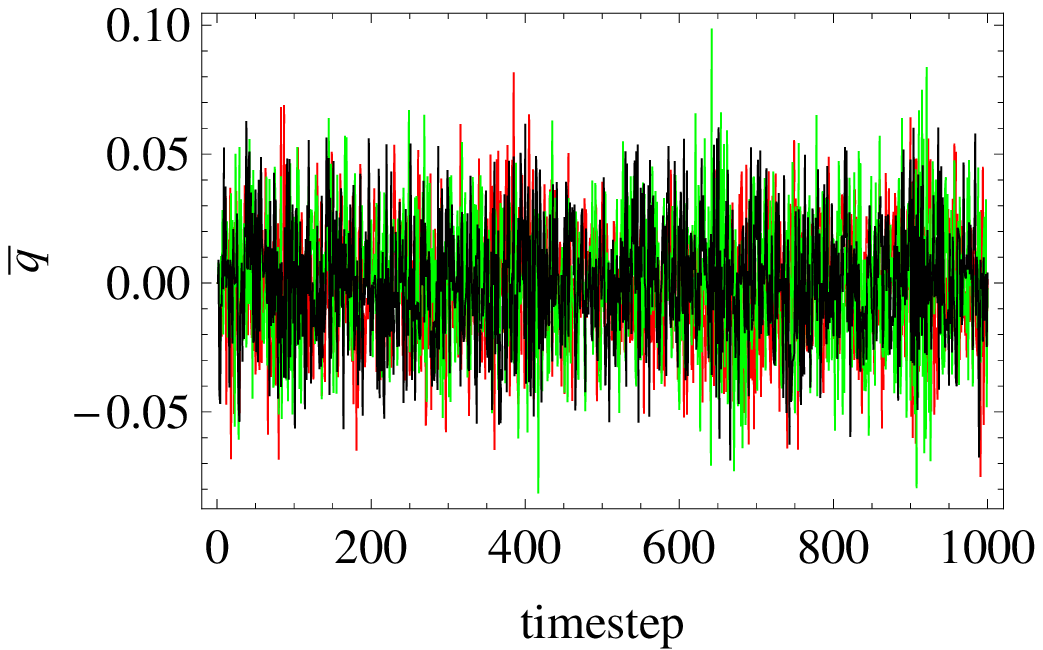}
  \includegraphics[width=8cm]{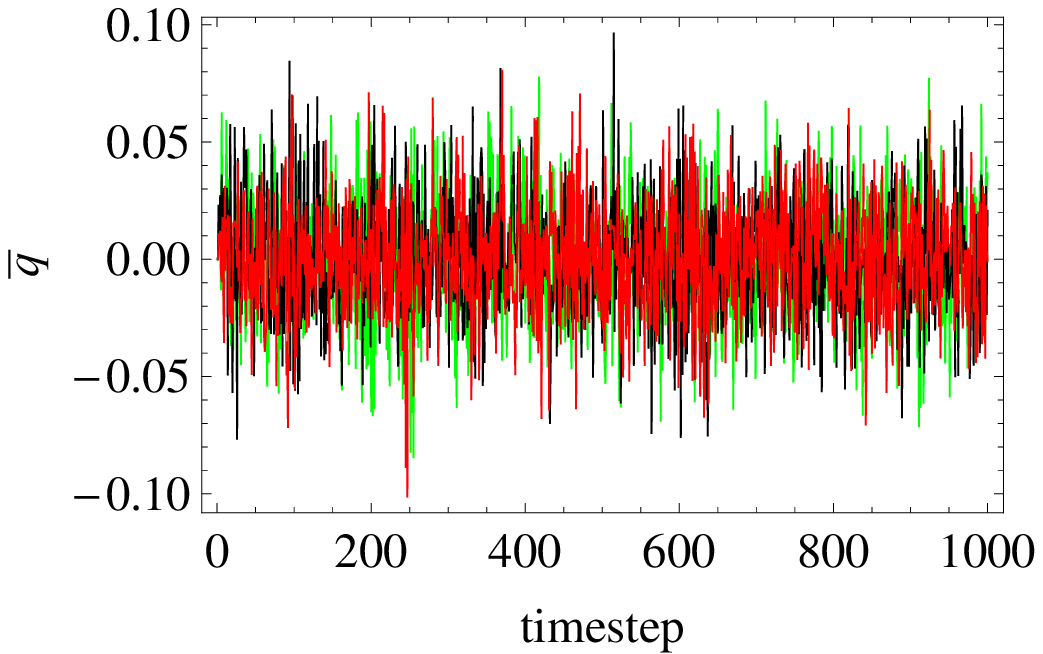}
   \caption{ Vertical displacement of the stochastically oscillating disk in the Schwarzschild geometry for different values of $a$ and $A$: $A = 0.0015$ and $a=0.03$ (red curve), $a = 0.1$ (green curve), $a=0.001$ (black curve) - left figure, and $a = 0.03$ and $A=1.5$ (red curve), $A = 0.15$ (green curve), and $A=0.015$ (black curve) - right figure.}
         \label{fig:sch-q1}
   \end{figure*}

      \begin{figure*}
   \centering
  \includegraphics[width=8cm]{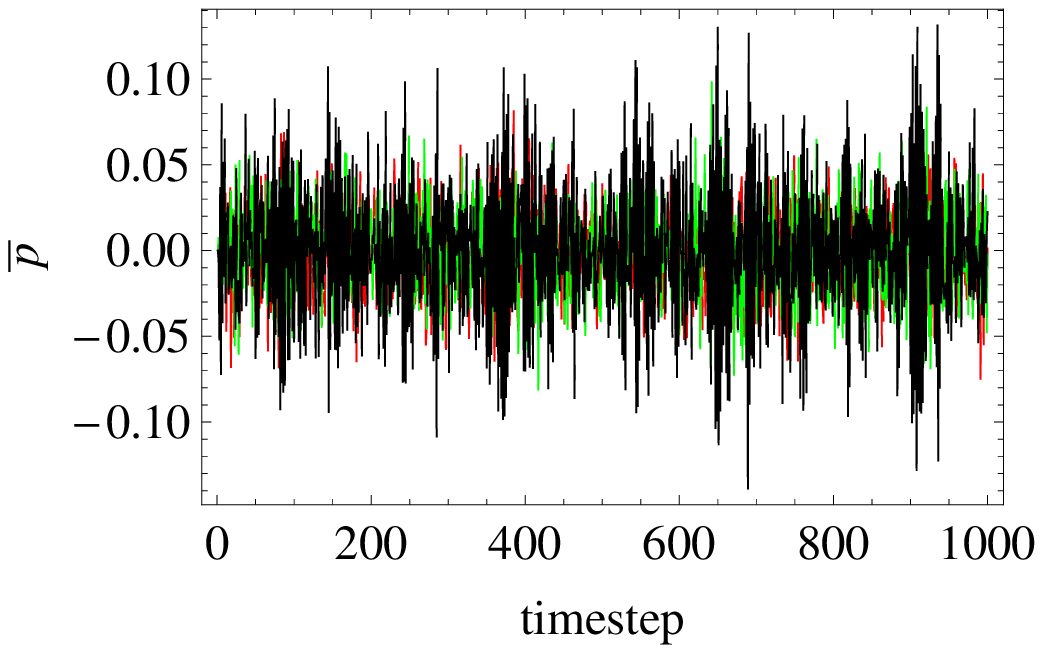}
   \includegraphics[width=8cm]{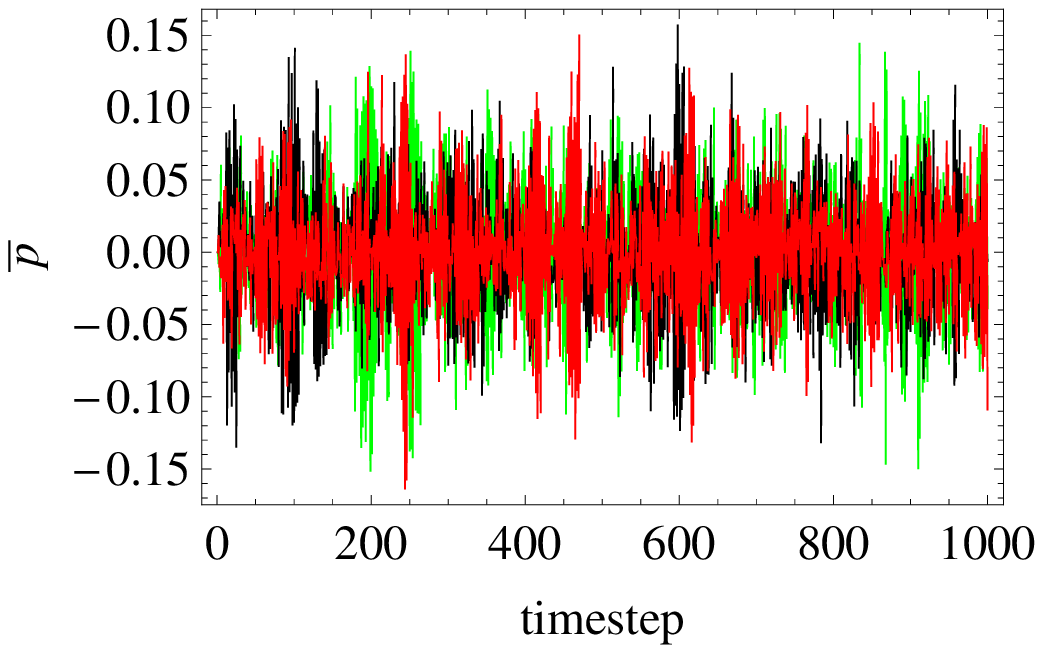}
   \caption{Velocity of the disk oscillations for the Schwarzschild geometry for different values of $a$ and $A$:  $A = 0.0015$, and $a=0.03$ (red curve), $a = 0.1$ (green curve), $a=0.001$ (black curve) - left figure, and  $a = 0.03$ and $A=1.5$ (red curve), $A = 0.15$ (green curve), $A=0.015$ (black curve) - right figure.}
         \label{fig:sch-p1}
   \end{figure*}

      \begin{figure*}
   \centering
  \includegraphics[width=8cm]{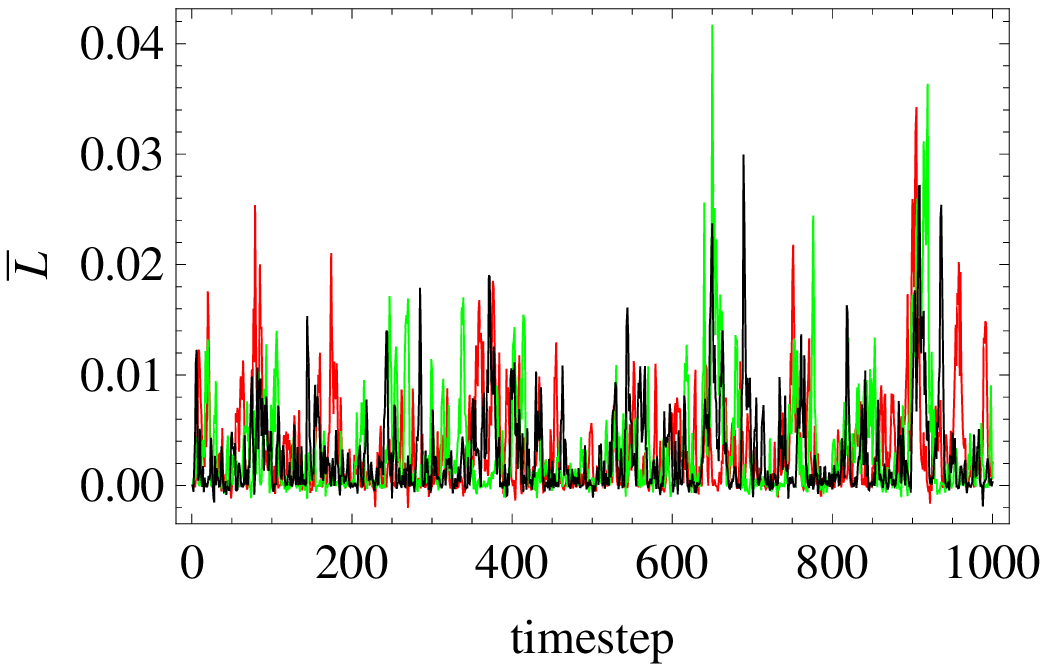}
  \includegraphics[width=8cm]{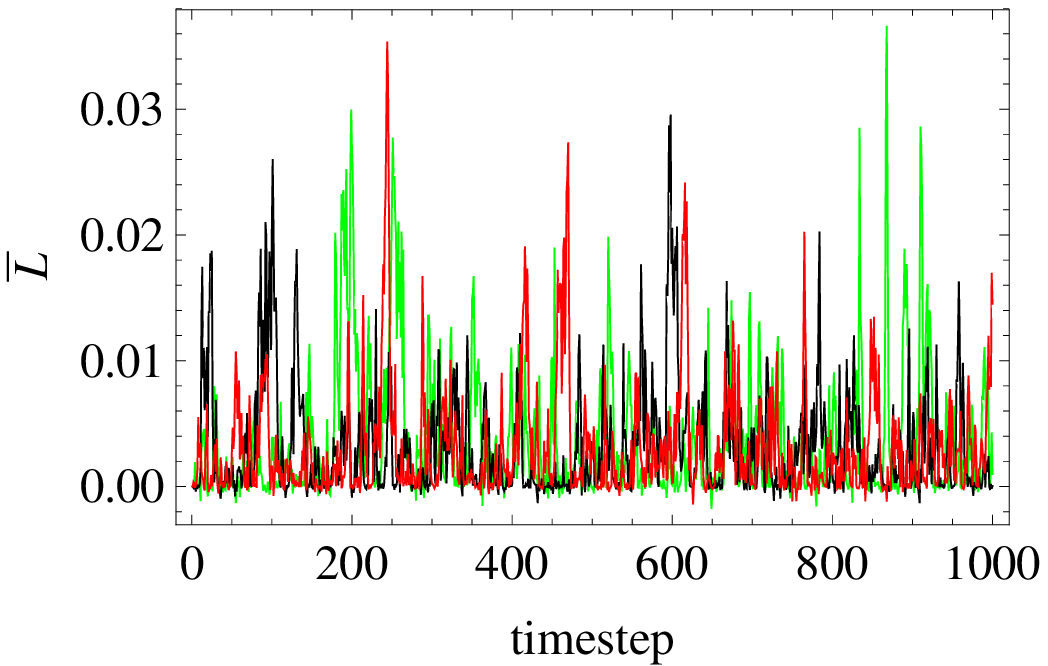}
   \caption{Luminosity of the stochastically oscillating disk in the Schwarzschild geometry, for different values of $a$ and A: $A = 0.0015$ and $a=0.03$ (red curve), $a = 0.1$ (green curve), $a=0.001$ (black curve)- left figure, and $a = 0.03$ and $A=1.5$ (red curve), $A = 0.15$ (green curve), $A=0.015$ (black curve) - right figure.}
         \label{fig:sch-lum1}
   \end{figure*}

The PSD of luminosity for the cases $a=0.03$, $A=0.0015$ and $a=0.03$ and $A=1.5$ are shown in Fig.~\ref{fig:sL1H2}, respectively. The PSD curves were obtained by using the .R software~\cite{Vaughan}.

  \begin{figure*}
   \centering
  \includegraphics[width=8cm]{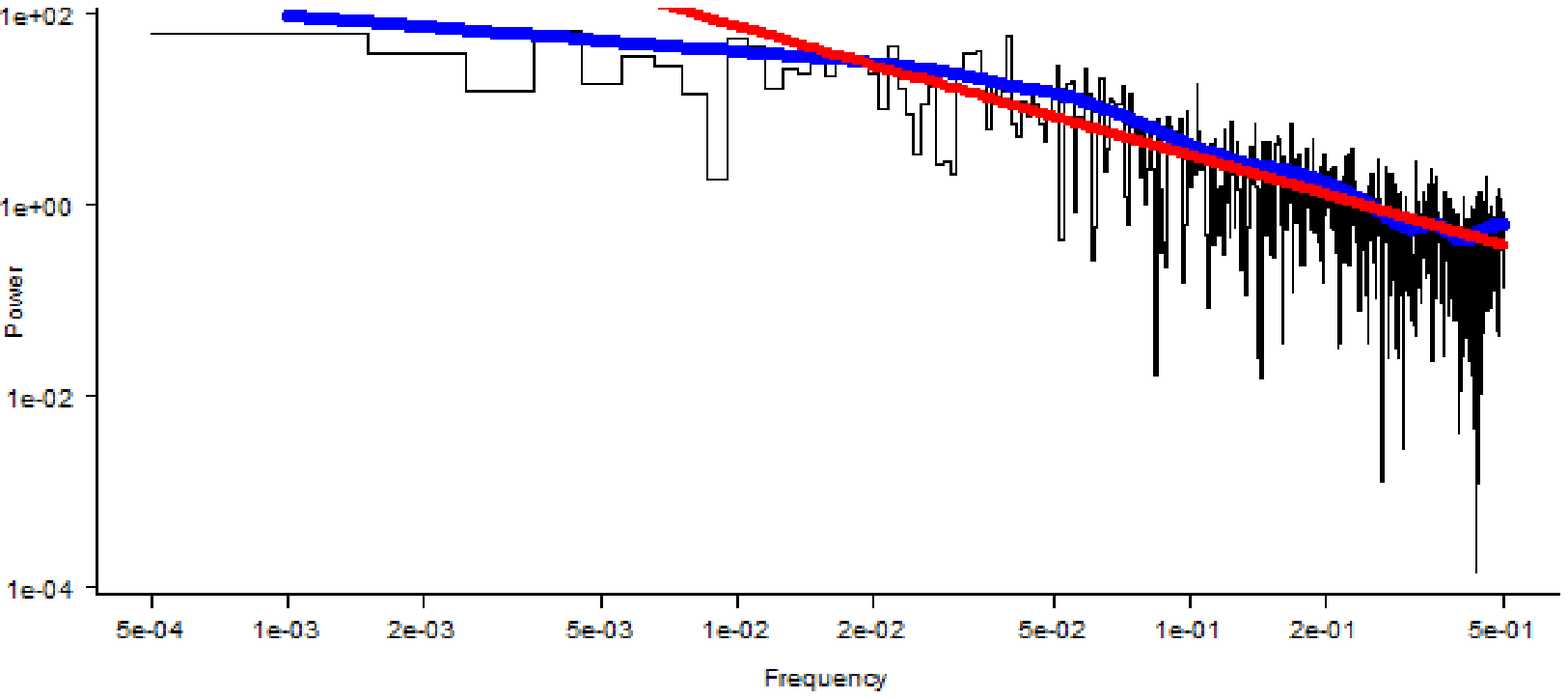}
   \includegraphics[width=8cm]{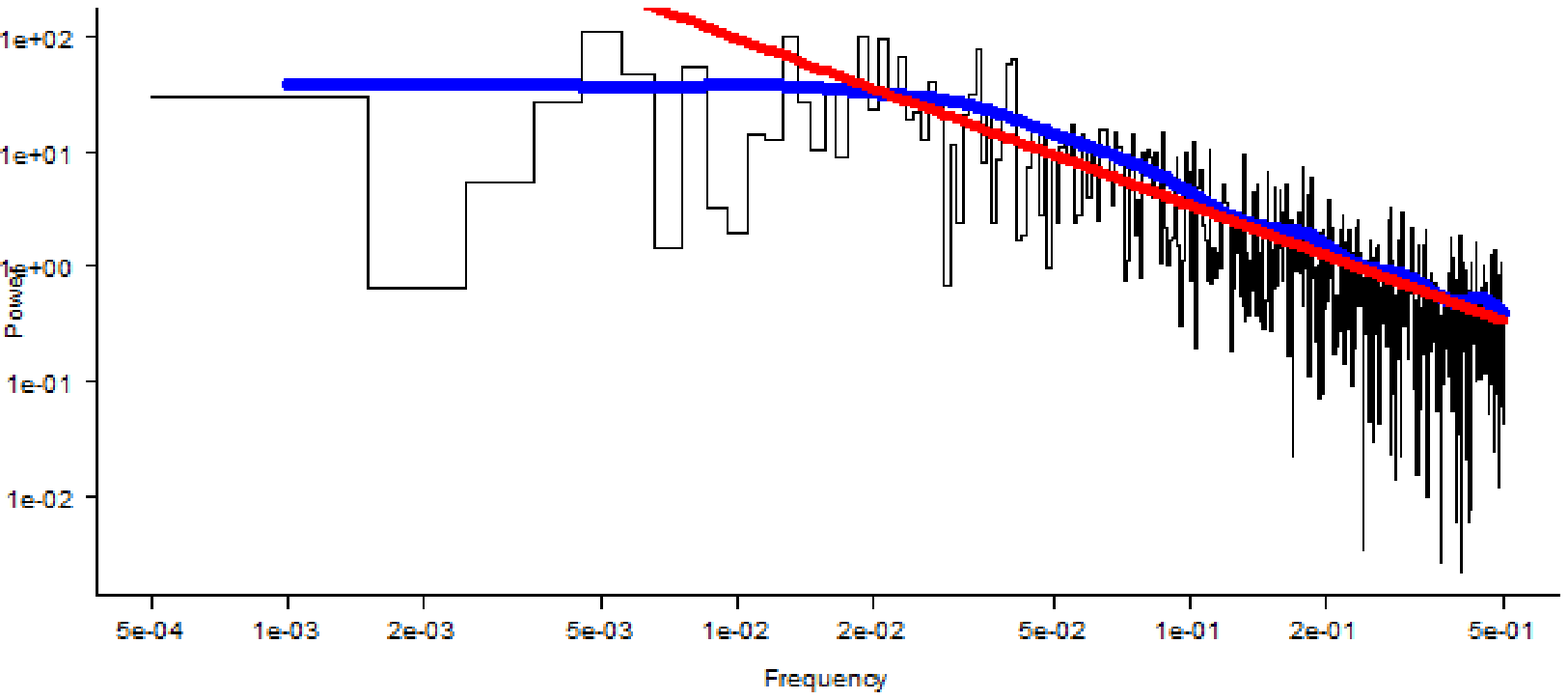}
   \caption{PSD of the luminosity for the Schwarzschild case,  for $a = 0.03$ and $A=0.0015$ (left figure). If a fit according to $P(f) \sim f^{\alpha}$ is attempted, $\alpha = -1.356$. PSD of luminosity for $a = 0.03$ and $A=1.5$ (right figure). If a fit according to $P(f) \sim f^{\alpha}$ is attempted, $\alpha = -1.455$.} \label{fig:sL1H2}
   \end{figure*}

   \subsection{Stochastic disk oscillations in the presence of colored noise in the Kerr geometry}

The line element of a stationary axisymmetric spacetime is given by the
Weyl-Lewis-Papapetrou metric~\cite{Kramer}
\begin{eqnarray}
ds^{2}&=&e^{2\left( \psi _1 -\psi _2 \right) }\left( d\rho ^{2}+dz^{2}\right)
+\rho ^{2}e^{-2\psi _2}d\phi ^{2}-\nonumber\\
&&e^{2\psi _2 }\left( cdt-\psi _3 d\phi \right)
^{2},
\end{eqnarray}
where $\psi _i$, $i=1,2,3$ are functions of $z$ and $\rho $ only. The
metric functions generating the Kerr black hole solution in axisymmetric
form are given by
\begin{equation}
\psi _1 =\frac{1}{2}\ln \frac{\left( R_{+}+R_{-}\right) ^{2}-4M^{2}+\left(
J^{2}/\vartheta ^{2}\right) \left( R_{+}-R_{-}\right) ^{2}}{4R_{+}R_{-}},
\end{equation}
\begin{equation}
\psi _2 =\frac{1}{2}\ln \frac{\left( R_{+}+R_{-}\right) ^{2}-4M^{2}+\left(
J^{2}/\vartheta ^{2}\right) \left( R_{+}-R_{-}\right) ^{2}}{\left(
R_{+}+R_{-}+2M\right) ^{2}+\left( J^{2}/\vartheta ^{2}\right) \left(
R_{+}-R_{-}\right) ^{2}},
\end{equation}
\begin{equation}
\psi _3 =-\frac{JM}{\vartheta ^{2}}\frac{\left( R_{+}+R_{-}+2M\right) \left[
\left( R_{+}-R_{-}\right) ^{2}-4\vartheta ^{2}\right] }{\left(
R_{+}+R_{-}\right) ^{2}-4M^{2}+\left( J^{2}/\vartheta ^{2}\right) \left(
R_{+}-R_{-}\right) ^{2}},
\end{equation}
where $M$ and $J$ are the mass and the specific angular momentum of the
compact object, $R_{\pm }=\sqrt{\rho ^{2}+\left( z\pm \vartheta \right) ^{2}}$, %
and $\vartheta =\sqrt{M^{2}-J^{2}}$.


By defining $\rho = nM$ as the radial distance from the origin in the plane of the disk, and the angular momentum of the compact object as $J = k M$, and denoting $\delta = M \sqrt{n^2+1-k^2}$, with $n>0$ and $k \in [0,1]$, the frequency of the free vertical oscillations of the general relativistic disks in the Kerr geometry  can be represented as \cite{Harko,Kato1}
\begin{equation}
\overline{\omega} _{\perp}^2=\frac{\omega _1^2\left(n,k\right)}{\omega _2^2\left(n,k\right)},
\end{equation}
where
\begin{eqnarray}
\omega _{1}^{2}\left( n,k\right)  &=&3n^{4}+n^{2}\sqrt{1+\delta }\times
\nonumber \\
&&\left[ 2k\left( 5+3\delta \right) +3\sqrt{1+\delta }\left( 6+3\delta
+\delta ^{2}\right) \right] - \nonumber\\
&&2\left( 1+\delta \right) ^{3/2}\times \Big[ -4k(2+\delta )+\nonumber\\
&&\sqrt{1+\delta }%
\left( -8+3\delta ^{2}+\delta ^{3}\right) \Big] ,
\end{eqnarray}
and
\begin{eqnarray}
\omega _2^2\left(n,k\right) &=& \left(1+\delta \right) \left [ n^2 - \delta \left(1+\delta
\right) \left(3+\delta \right) \right ]^2 \times \Bigg \{
\delta-\nonumber\\
&&  \frac{\left [ k+ \left(1+\delta \right)^{3/2}\right
]^2 \left [ 4 \left(1+\delta \right) + n^2 \left(3+\delta
\right)\right]}{\left [ n^2 - \delta \left(1+\delta \right)
\left(3+\delta \right)\right]^2}+ \nonumber\\
&&\frac{4k\left[k +
\left(1+\delta \right)^{3/2}\right]}{n^2-\delta \left(1+\delta
\right) \left(3+\delta \right)}-1\Bigg\},
\end{eqnarray}
respectively.

By adopting for the mass of the central object a value of $M=10^8M_\odot$, and by fixing $n = 10$, $k =0.9$, and $\overline{\omega}_\perp ^2 = 0.0012$, respectively,  and considering for the step and the initial conditions the values $h=1$, $\overline{q}(0) = 0$, $\overline{p} (0) = 0.01$, $b=4$, we have analysed two cases. All trajectories are obtained by mediating over 1000 stochastic trajectories (for which the variable $\overline{z}_0$ is redrawn from its distribution).

For fixed $A = 0.0015$ we varied $a$ and obtained the behaviour of the dimensionless displacement, dimensionless velocity, and dimensionless luminosity. Then, for $a=0.03$ we varied $A$ and obtained the behaviour of the same physical parameters of the disk. The results of the numerical simulations are presented in Figs.~\ref{fig:kerr-q1}, \ref{fig:kerr-p1} and \ref{fig:kerr-lum1}, respectively.

   \begin{figure*}
   \centering
  \includegraphics[width=8cm]{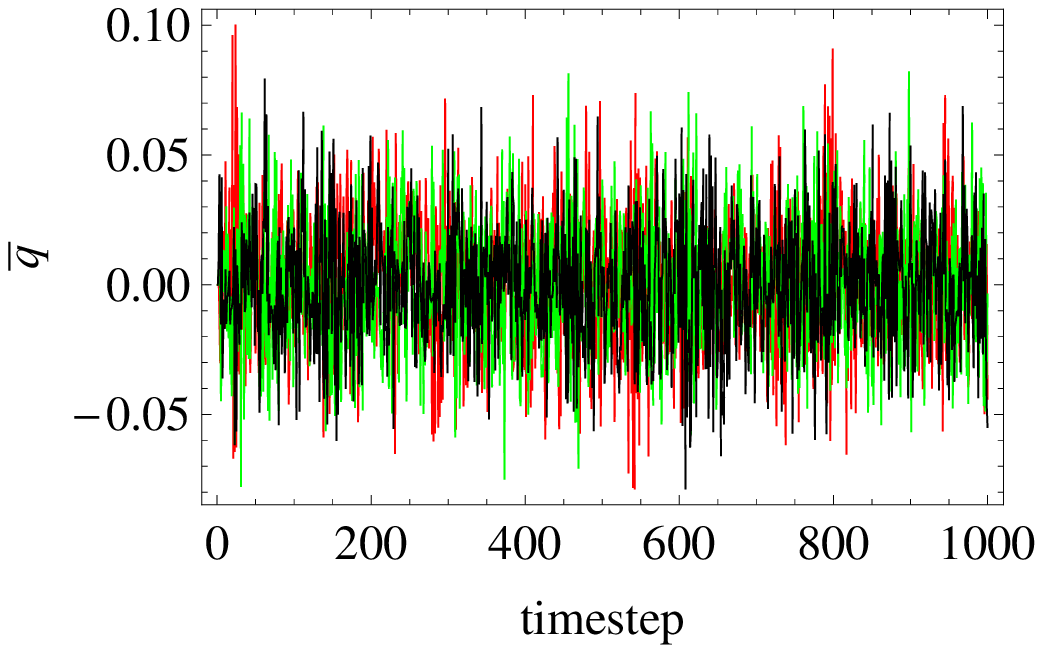}
   \includegraphics[width=8cm]{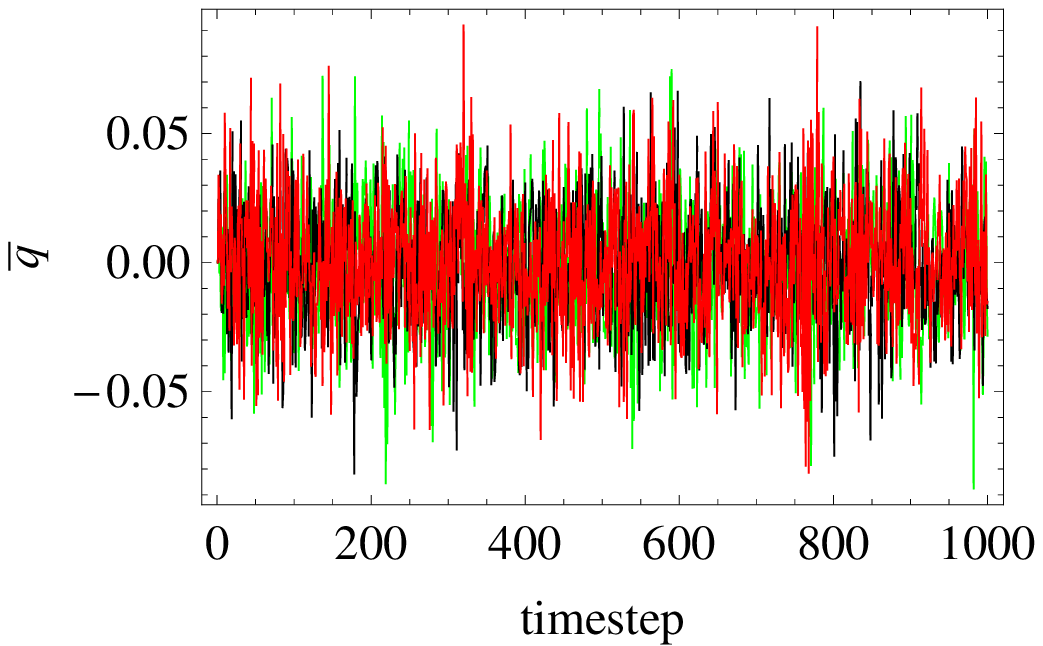}
   \caption{Vertical displacement of the stochastically oscillating disk in the Kerr geometry for $A = 0.0015$ and $a=0.03$ (red curve), $a = 0.1$ (green curve), $a=0.001$ (black curve) - left figure, and for $a = 0.03$ and $A=1.5$ (red curve), $A = 0.15$ (green curve), $A=0.015$ (black curve) - right figure.}
         \label{fig:kerr-q1}
   \end{figure*}

      \begin{figure*}
   \centering
  \includegraphics[width=8cm]{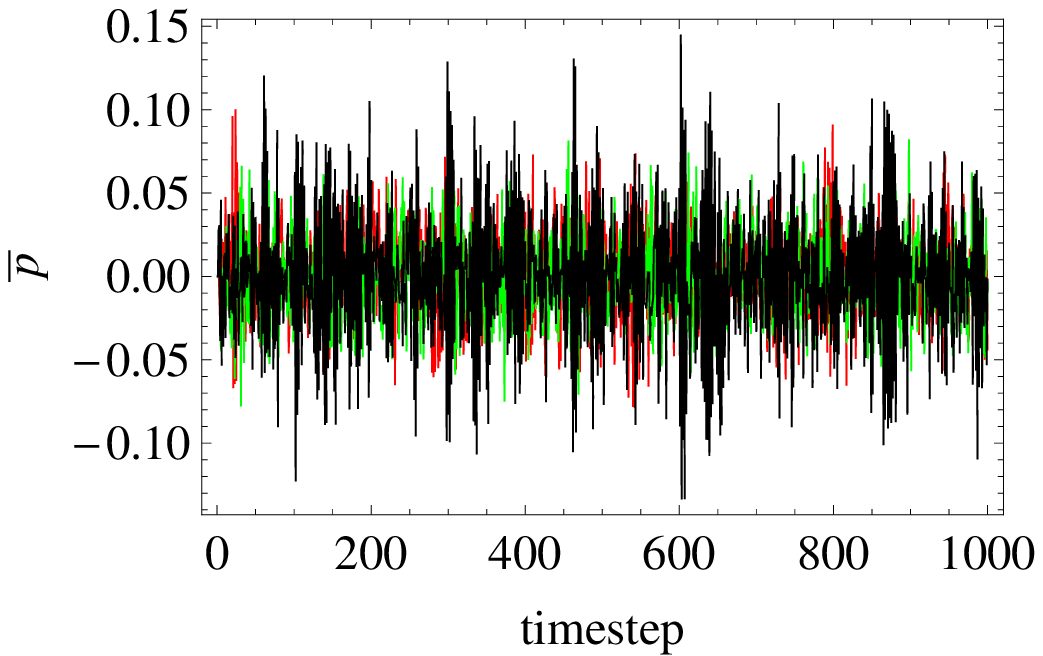}
  \includegraphics[width=8cm]{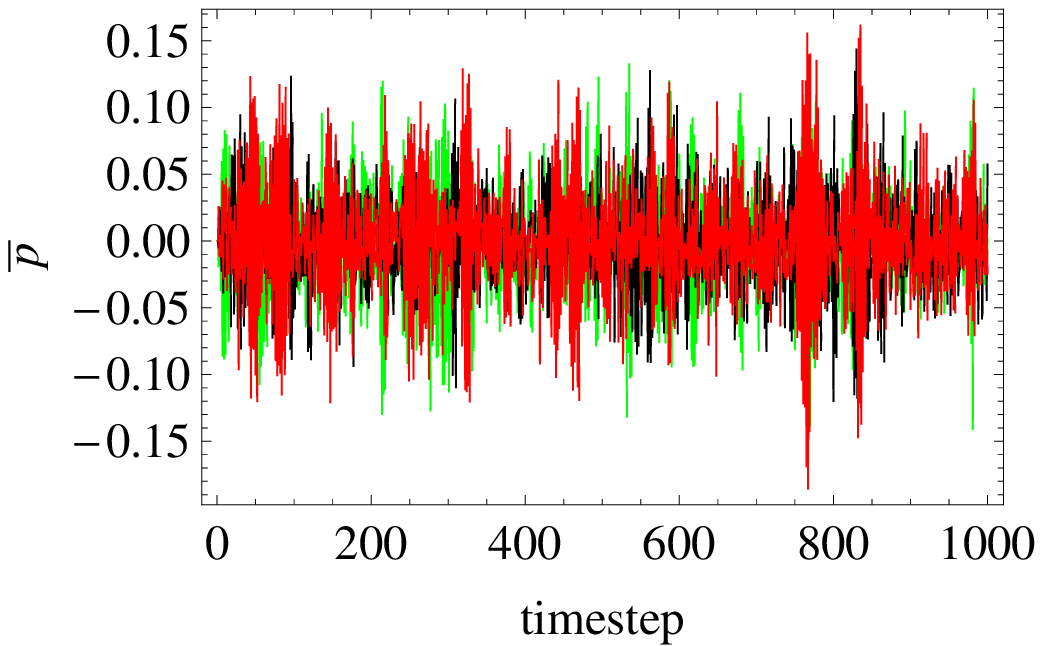}
   \caption{Velocity of the stochastically oscillating disk in the Kerr geometry, for $A = 0.0015$ and $a=0.03$ (red curve), $a = 0.1$ (green curve), $a=0.001$ (black curve) - left figure, and for $a = 0.03$ and $A=1.5$ (red curve), $A = 0.15$ (green curve), $A=0.015$ (black curve) - right figure.}
         \label{fig:kerr-p1}
   \end{figure*}

      \begin{figure*}
   \centering
  \includegraphics[width=8cm]{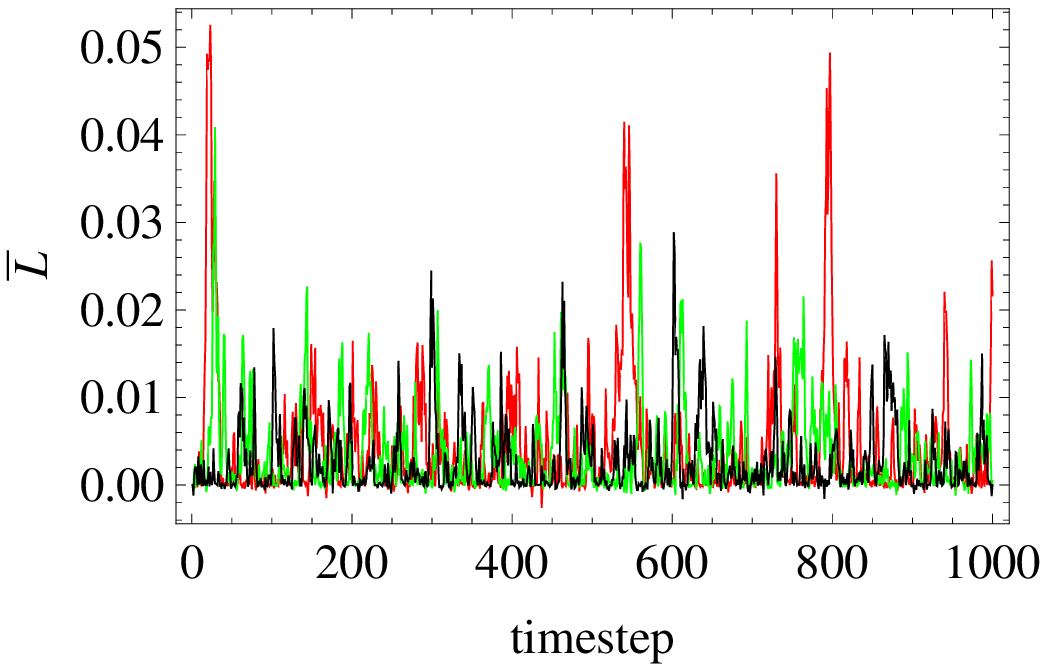}
   \includegraphics[width=8cm]{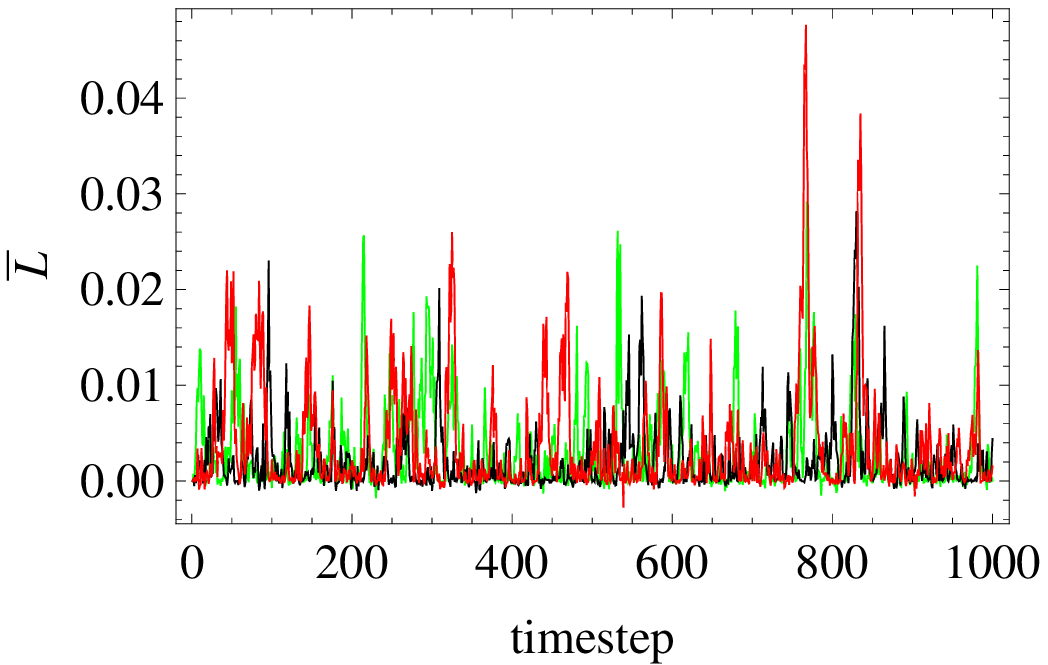}
   \caption{The  luminosity of the stochastically oscillating disk in the Kerr geometry for $A = 0.0015$ and $a=0.03$ (red curve), $a = 0.1$ (green curve), $a=0.001$ (black curve) - left figure, and for $a = 0.03$ and $A=1.5$ (red curve), $A = 0.15$ (green curve), $A=0.015$ (black curve) - right figure.}
         \label{fig:kerr-lum1}
   \end{figure*}

The PSDs of the luminosity for the cases $a=0.03$, $A=0.0015$ and $a=0.03$, $A=1.5$, respectively, are shown in Fig.~\ref{fig:kL1H2}. Values for the spectral slope obtained for $a=0.03$, $A=0.0015$ and different configurations of the system producing the simulated light curves are given in Table~\ref{table:simAlpha}. The fourth column in Table~\ref{table:simAlpha} is the Bayesian probability that the source actually behaves like in the attempted fit, where $p_B$ close to $1$ suggests that the fit is correct. The definition and calculation of $p_B$ is quite lengthy, but we sketch it here and refer the reader to~\cite{Vaughan} for full detail. In cases such as investigating data from an AGN, when there is only one light curve available (call it $\vec{x}^{obs}$), where reproducibility of the experiment is not an option, the .R software with the Bayes script offers the possibility to asses whether or not the recorded data was produced by a process behaving like a null hypothesis, $H_0$. For this purpose, the script simulates numerous sets of data based on the statistical properties of $\vec{x}^{obs}$, stores them in $\vec{x}^{rep}$ and treats these new data sets as they would be multiple measurements of the same process. Afterwards it defines $\Xi(\vec{x})$, called a test statistics, which is a real valued function of the data chosen such that extreme values are unlikely when the null hypothesis $H_0$ is true. With these parameters, the Bayesian probability is defined as
\begin{equation}
p_B (\vec{x}) = \text{Pr} \left \{ \Xi(\vec{x}^{rep}) \leq \Xi(\vec{x}^{obs}) | \vec{x}^{obs}, H_0 \right \},
\end{equation}
where $\text{Pr} = \{ x | y \}$ means probability of event $x$ given that event $y$ has occurred.

  \begin{figure*}
   \centering
  \includegraphics[width=8cm]{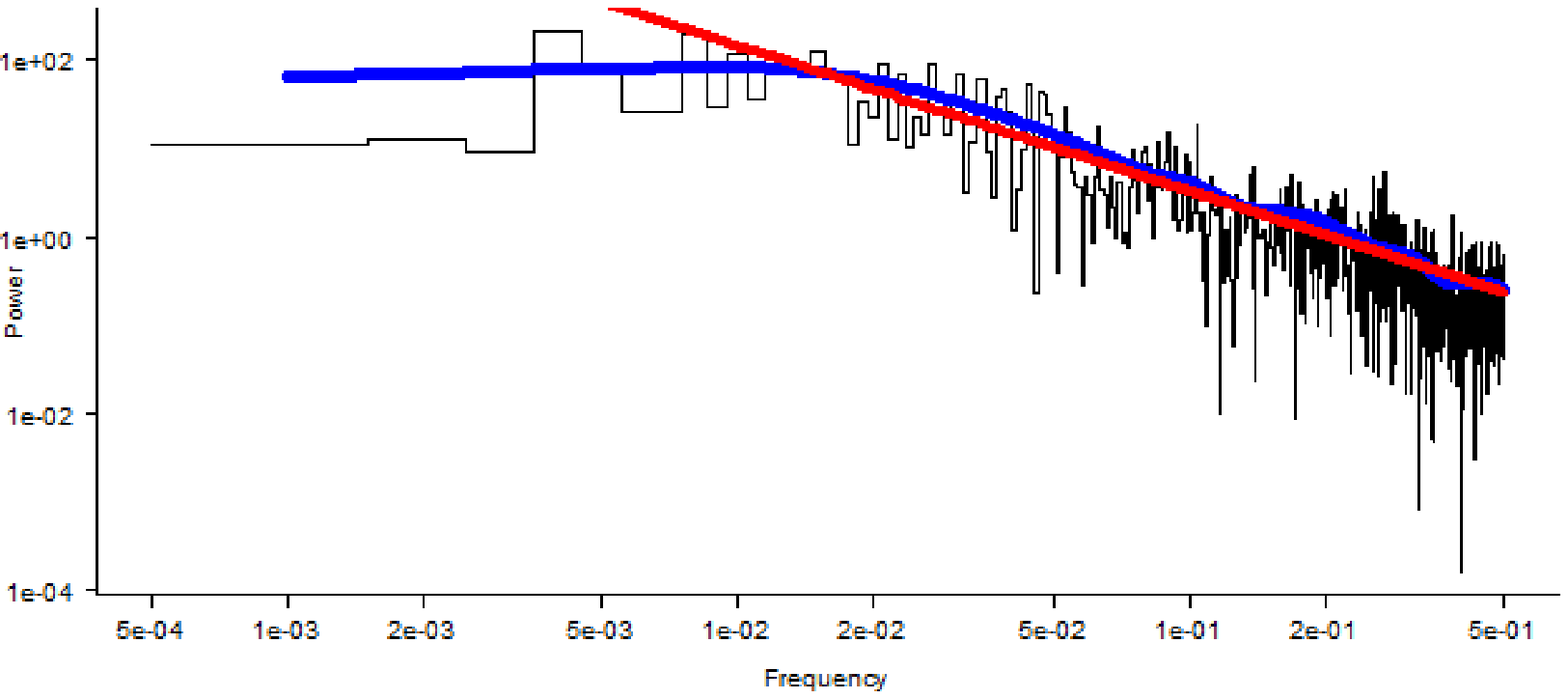}
   \includegraphics[width=8cm]{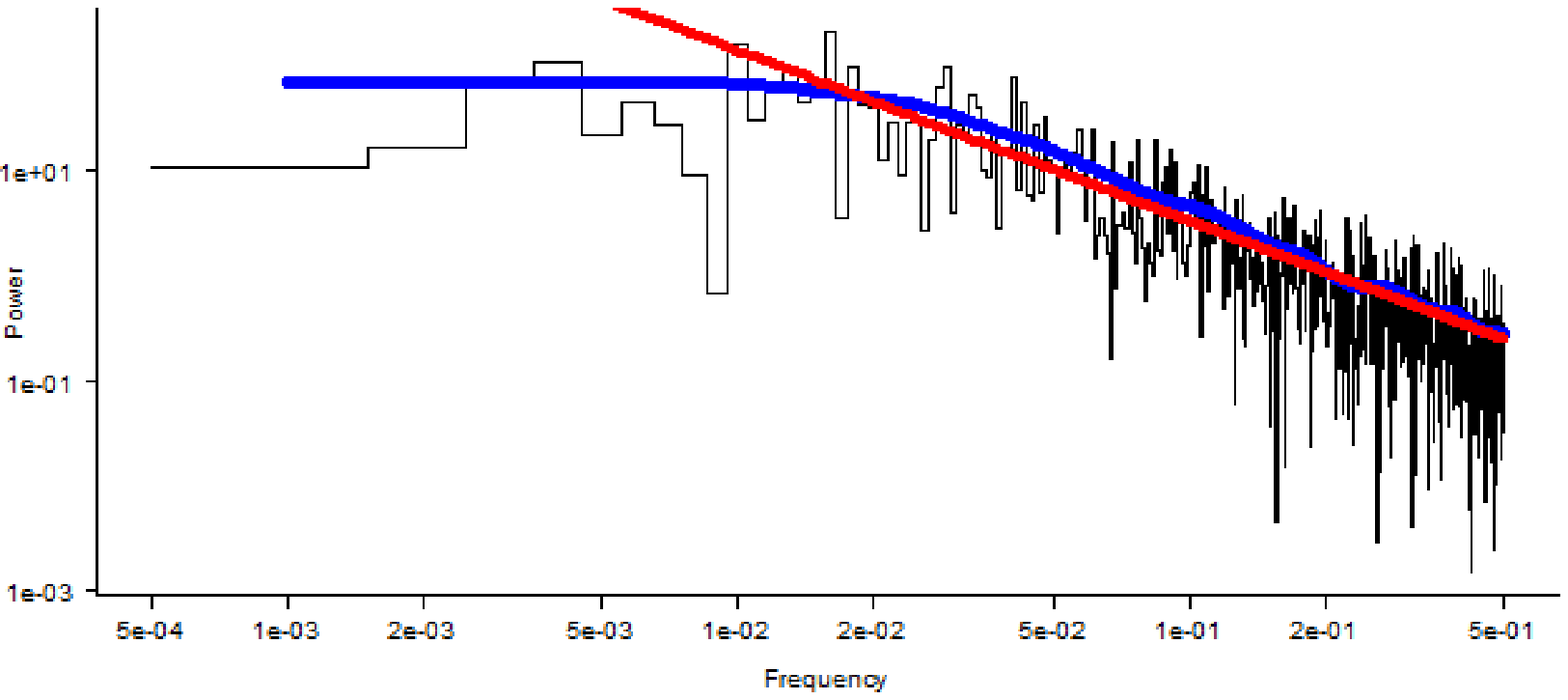}
   \caption{ PSD's of the luminosity of the stochastically oscillating disk in the Kerr geometry. In the left figure,  $a = 0.03$ and $A=0.0015$  If a fit according to $P(f) \sim f^{\alpha}$ is performed, $\alpha = - 1.65$. In the right figure $a = 0.03$ and $A=1.5$. If a fit according to $P(f) \sim f^{\alpha}$ is performed, $\alpha = -1.619$.}
         \label{fig:kL1H2}
   \end{figure*}

   \begin{table}
   \begin{center}
\begin{tabular}{|l|c|c|c|}
\hline
  k   &  n  & $-\alpha$          & $p_B$ \\  \hline
  0   &  10 & 1.356 [$\pm$ 0.05] & 0.749 \\ \hline
  0.9 &  10 & 1.650 [$\pm$ 0.05] & 0.064 \\ \hline
  0.5 &   7 & 1.480 [$\pm$ 0.05] & 0.288 \\ \hline
  0.5 &   8 & 1.372 [$\pm$ 0.05] & 0.755 \\ \hline
  0.5 &   9 & 1.499 [$\pm$ 0.05] & 0.717 \\ \hline
  0.9 &   7 & 1.387 [$\pm$ 0.05] & 0.289 \\ \hline
  0.9 &   8 & 1.369 [$\pm$ 0.05] & 0.079 \\ \hline
  0.9 &   9 & 1.562 [$\pm$ 0.09] & 0.087 \\ \hline
  \end{tabular}
  \caption{Values for the spectral slope of the luminosity curves, obtained by using the .R software, for $a=0.03$, $A=0.0015$, and different mass and angular momentum configurations of the black hole - accretion disk system producing the simulated light curves. The values given in the column for $-\alpha$ represent the mean values of the parameter calculated from the series simulated with .R (as explained in the text), followed by $[\pm \text{ standard deviation}]$.}\label{table:simAlpha}
  \end{center}
\end{table}

  \section{Comparison with AGN IDV data}\label{sect5}

Extensive observational and theoretical efforts have been made in
order to explain IntraDay Variability (IDV) in some classes of
Active Galactic Nuclei (AGN)~\cite{Krichbaum,Wagner,Poon,Min,Balbus}. IDV manifests as the fast change (in less than one day) in the luminosity output of an object, and for supermassive black holes IDV occurs in the optical domain. The PSD of these light curves are found to be nontrivial~\cite{Carini,Azarnia,Leung2011}.

 In the present paper we assume that the IDV emission occurs from the disk of a super-massive black hole - accretion disk system \cite{Min,Leung}, and  it is due to the  oscillations of the disk perturbed by a sudden interaction with the external environment. Such a transient interaction can be due to the stochastic gravitational force generated by the fluctuations of the star number in the near neighborhood of the black hole - accretion disk system, if the system is located in a dense stellar cluster,  or to the interaction of the disk  with the debris produced by the tidal disruption of a star captured by the central black hole.

For BL Lac S5 0716+714, an AGN with  $M \in [10^{7.68}M_\odot,$ $ 10^{8.38}M_\odot]$~\cite{Fan}, a PSD analysis performed on IDV data in the BVRI bands has shown that $\alpha$ varied between approximately $-1.5$ and $-2.7$ for that set of data \cite{Mocanu1}.

There are indications that the relationship between the root-mean-square-deviation (rms) and the flux provides more information regarding the source-process of fast variability, at least in X-Ray Binaries~\cite{Uttley}. For these objects, for which IDV is exhibited in the X-Ray domain, it was found that this relationship is linear. However, a test performed on optical IDV for BL Lac S5 0716+714 on observational data from different epochs showed that the rms-flux relation was not linear~\cite{Mocanu2}.

In order to test if the optical IDV flux data of BL Lac S5 0716+714 are log-normally distributed,  in Fig.~\ref{fig:lognormData} we compare the observational data with the normal distribution. The bins represent the logarithm of the observational data, and the smooth line is a superimposed Gaussian with the same mean and variance as the data. If the two would have coincided (i.e. if the flux would have been log-normally distributed), the rms-flux relation would have been linear. We find that the simulated data, obtained from the generalized Langevin equation description of the luminosity of the stochastically oscillating disks, also exhibit this feature, i.e., that the rms-flux relation is not linear, for neither of the Schwarzschild or the Kerr case (an example is shown in Figs.~\ref{fig:lognormSim}).

  \begin{figure}
   \centering
  \includegraphics[width=8cm]{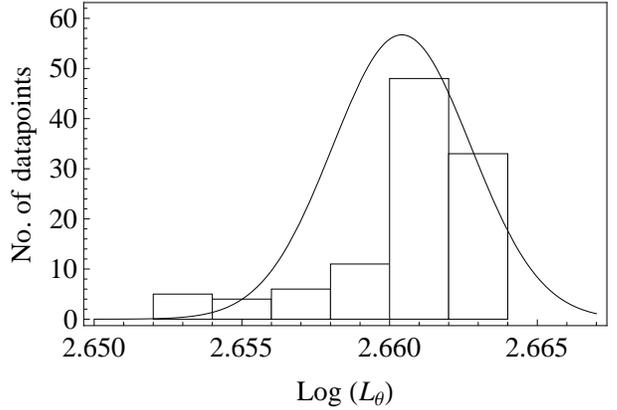}
   \caption{Test of the log-normal distribution of the optical IDV flux data of BL Lac S5 0716+714 \cite{Poon}. If the bins (the distribution of the data) can be fitted by the continuous curve (a normal distribution with the same mean and variance as the data), the distribution of the data is log-normal. The data was recorded in the B filter, on JD 2454824.}
         \label{fig:lognormData}
   \end{figure}

  \begin{figure}
   \centering
  \includegraphics[width=8cm]{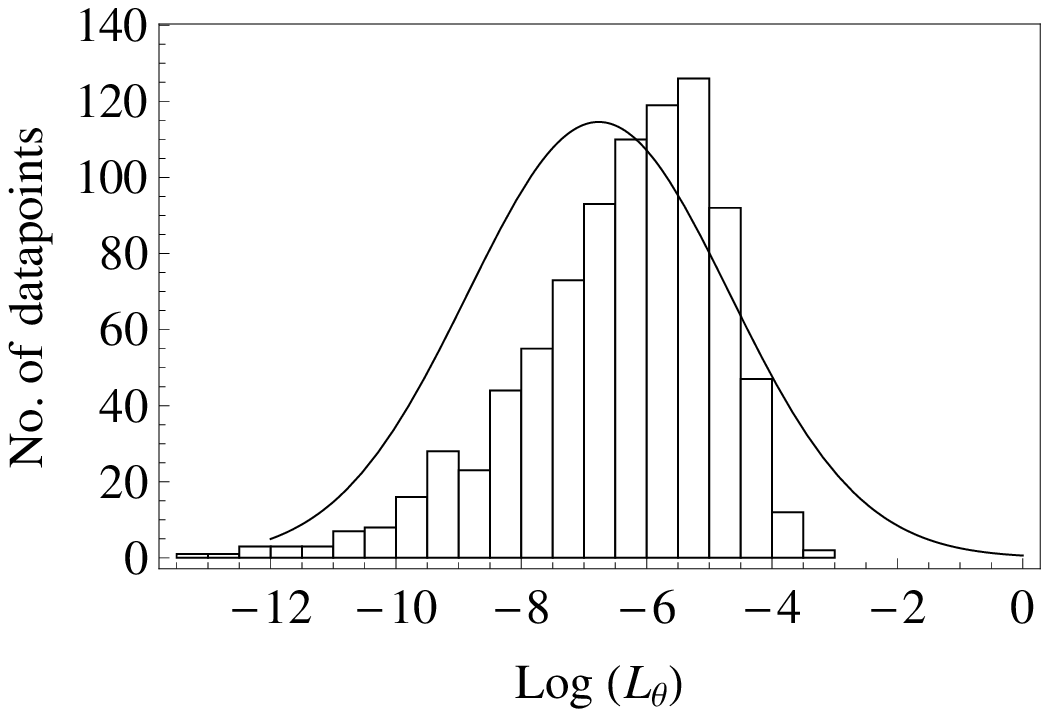}
   \includegraphics[width=8cm]{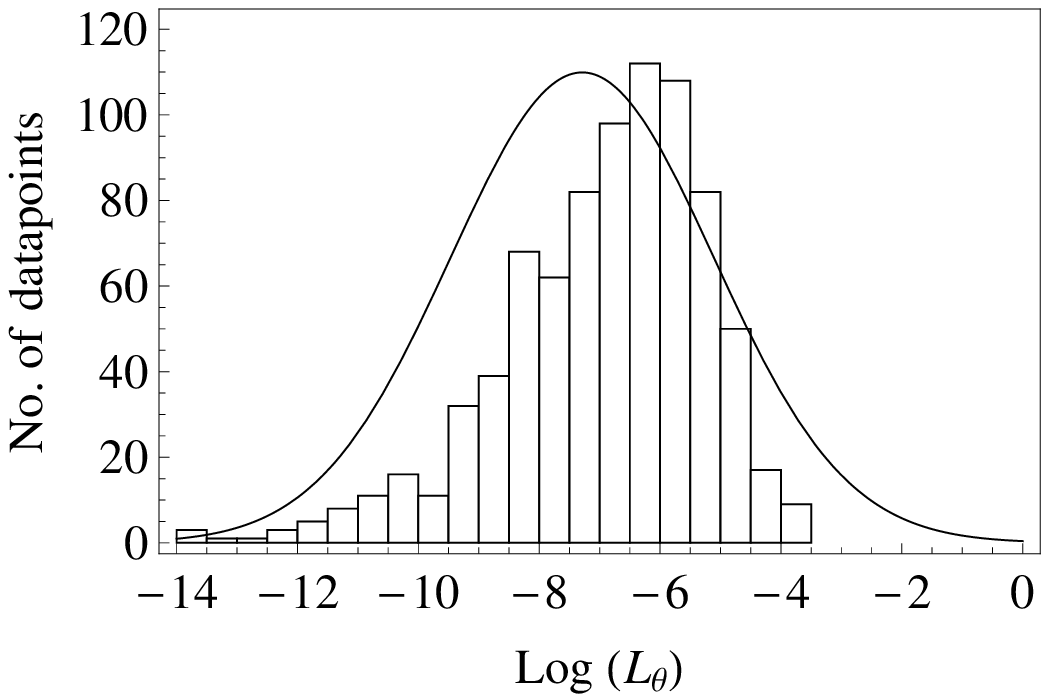}
   \caption{Test of the log-normal distribution of the light curves generated by the stochastically oscillating accretion disks. The simulated data are represented by the bins. The continuous superimposed line is a normal distribution with the same mean and variance as the simulated data. Left panel: Schwarzschild case, with $A=0.0015$, $a=0.03$ and $n=10$. Right panel: Kerr case, with $A=0.015$, $a=0.03$, $n=10$ and $k=0.9$.}
         \label{fig:lognormSim}
   \end{figure}

The model presented in this work is thus consistent with observational data, from the point of the statistics exhibited by the light curves. This is very encouraging and supports the idea that perturbations due to colored/nontrivially correlated noise consistently explain the statistical properties of the observed IDV light curves.

\section{Discussions and final remarks}\label{sect6}

In the present paper we have introduced a mathematical description of the stochastic vertical oscillations of the accretion disks based on the generalized Langevin equation with colored noise and memory effects. From an astrophysical point of view this approach may be relevant for modeling of accretion disks around compact general relativistic objects randomly interacting with the cosmic environment surrounding the disk.  Events like the encounter of a star, or a massive compact object, by the central black hole, or the interaction between the disk and the cosmic dust resulted from the tidal disruption of a star captured by the central black hole, may trigger such interactions, which are essentially random in their nature. In our description we have assumed that the oscillation velocities of the massive test particles composing the disk are small, and therefore they can be described by the Newtonian equation of motion. Hence in our approach we have assumed that the motion of the disk is non-relativistic in the sense that the velocity $\delta z/dt$ of the particles is much smaller than the speed of light, $\delta z/dt<<c$. However, since the gravitational field in the  vicinity of the black hole is strong, we have adopted for the oscillation frequency of the disk in the gravitational field the general relativistic values. The cases of the static and rotating black holes were considered independently, and the effect of the rotation on the disk luminosity was explicitly considered. The analysis of the PSD of the luminosity curves has shown that in the presence of memory effects and colored noise non-trivial values can be obtained for the spectral slope. Therefore the presence of the colored noise/memory effects could allow a better description of the astrophysical processes involving random/stochastic factors.

 In the present paper we have considered that accretion disks are composed from massive test particles forming a fluid system, and  moving on non - geodesic lines around the central black hole. By considering the equation of motions of the disk perturbations, and assuming that the gravitational effects dominates the hydrodynamic ones, in the first order of approximation the effects of the non-zero pressure and energy density give in the equation of motion a correction term proportional to the magnitude of the vertical perturbations $\delta z$.  In the non-relativistic limit the disk pressure effects and the external gravitational perturbation terms can be combined in a single, effective coefficient, proportional to the vertical displacement, and which determines the frequency of the disk oscillations.   In order to construct realistic physical models one should assume an equation of state for the disk matter components, and adopt some particular disk structures. Since there is no general consensus on the physical properties of the accretion disks, there is no guarantee that adopting a particular disk model would produce observationally relevant results. On the other hand, in the presence of a strong gravitational field, the gravitational effects dominate the pressure effects, which can be neglected with a very good approximation. In this case the motion of the particles in the disk can be considered as geodesic. This approximation is certainly valid for particles located in the inner region of the disk. As for the viscous dissipative effects, in the present paper we have assumed that they can be described by a single integral kernel term, whose functional form has been fixed phenomenologically, so that the fluctuation - dissipation theorems can be satisfied. However, some essential features of the accretion disks dynamics can be obtained, and analyzed, even by the present simplified theoretical model.

As a possible astrophysical application of our model we have compared a set of  IDV observational light curve data with the results of the simulations of the vertical stochastic oscillations of the accretion disks, described by the generalized Langevin equation. Some common features of the theoretical model and observations, namely, the possibility of obtaining non-trivial PSD values, and the deviation in both cases of data from a log-normal distribution, were identified. Hence the generalized Langevin equation with colored noise could provide a powerful tool in the description of the physical processes  in the vicinity of compact general relativistic objects.

\section*{Acknowledgments}

We would like to thank to the anonymous referee for comments and suggestions that helped us to significantly improve our manuscript. GM is supported by a grant of the Romanian National Authority of Scientific Research, Program for Research - Space Technology and Advanced Research - STAR, project number 72/29.11.2013.

\end{document}